\documentclass[entropy,accept,moreauthors,pdftex]{Definitions/mdpi} 
\firstpage{1} 
\pubvolume{23}
\issuenum{2}
\articlenumber{230}
\pubyear{2021}
\copyrightyear{2021}
\externaleditor{Academic Editor: J\"{u}rgen Pilz} % For journal Automation, please change Academic Editor to "Communicated by"
\datereceived{20 January 2021} 
\dateaccepted{11 February 2021} 
\datepublished{16 February 2021} 
\hreflink{https://\linebreak doi.org/10.3390/e23020230} % Please use \linebreak if need to do line break
\usepackage{subfigure}
\makeatletter
\renewcommand{\@thesubfigure}{\normalsize(\textbf{\alph{subfigure}})}
\makeatother

\usepackage{cases}
\usepackage{natbib}
\usepackage{subfigure}
\usepackage{comment}
\newcommand{\R}{\mathbb{R}} % real line
\newcommand{\Rp}{\R^{p}} % p-dimensional real vector space
\newcommand{\Rnp}{\R^{n\times p}} % n*p real matrix space
\newcommand{\Rpp}{\R^{p\times p}} % p*p real matrix space
 
\newcommand{\inv}{^{-1}} % inverse 
\newcommand{\tp}{^{\top}} % transpose 

\newcommand{\E}{\mathbb{E}} % expectation

\newcommand{\labelvector}[1]{\boldsymbol{#1}} % use additional {} if bm package is applied
\newcommand{\labeltarget}[1]{#1} % target---unmodified
\newcommand{\labelestimator}[1]{\widehat{#1}} % estimator

\newcommand{\bfl}{[}%
\newcommand{\bfr}{]}%

\newcommand{\bflb}{\bigl[}%
\newcommand{\bfrb}{\bigr]}%

\newcommand{\bflbb}{\Bigl[}%
\newcommand{\bfrbb}{\Bigr]}%

\newcommand{\bflbbbb}{\Biggl[}%
\newcommand{\bfrbbbb}{\Biggr]}%

\newcommand{\bslu}{\{}% % for subscripts
\newcommand{\bsru}{\}}%

\newcommand{\bslb}{\bigl\{\mskip 2mu}%
\newcommand{\bsrb}{\mskip 2mu\bigr\}}%

\newcommand{\bslbbb}{\biggl\{\mskip 2mu}%
\newcommand{\bsrbbb}{\mskip 2mu\biggr\}}%

\newcommand{\norm}[1]{\ensuremath{|\!| {#1} |\!|}} % norm notation
\newcommand{\normtwos}[1]{\ensuremath{\norm{#1}_2^2}} % square of l2-norm

\newcommand{\design}{\boldsymbol{X}}
\newcommand{\designrow}{\labelvector{x}}

\newcommand{\outcomes}{y}
\newcommand{\outcome}{\labelvector{\outcomes}}
\newcommand{\targets}{\labeltarget{\beta}}
\newcommand{\target}{\boldsymbol{\labeltarget{\targets}}}

\newcommand{\estimator}{\labelestimator{\target}}
\newcommand{\noises}{u}
\newcommand{\noise}{\labelvector{\noises}}
\newcommand{\variance}{\sigma^2}

\newcommand{\CovMatrix}{\boldsymbol{\Sigma}}

\newcommand{\penaltycoef}{\tau}
\newcommand{\penaltyfun}{h_{\penaltycoef}}
\newcommand{\regressors}{\alpha}
\newcommand{\regressor}{\labelvector{\regressors}}

\newcommand{\numvar}{p}  % number of variables
\newcommand{\samplesize}{n}
\newcommand{\loglikfuc}{l}

\newcommand{\argmin}{\arg\min}

\newcommand{\zero}{\boldsymbol{0}}
\newcommand{\identitymatrix}{\boldsymbol{I}}

\newcommand{\numknockoffs}{k}
\newcommand{\numFDR}{r}
\newcommand{\Wstatistics}{W}
\newcommand{\threshold}{T}

\newcommand{\Zstatistics}{Z}
\newcommand{\Zknockoffs}{\widetilde{Z}}
\newcommand{\designknockoffs}{\widetilde{\boldsymbol{X}}}
\newcommand{\designrowknockoffs}{\widetilde{\designrow}}
\newcommand{\diagv}{\labelvector{a}}

\newcommand{\trueas}{\mathcal S^*} % true active set
\newcommand{\estas}{\widehat{\mathcal S}} % estimated active set
\newcommand{\estasq}{\estas_\fdr} % estimated active set depends on q
\newcommand{\estasqone}{\estas_{\fdr_1}}
\newcommand{\estasqi}{\estas_{\fdr_i}}
\newcommand{\estasqk}{\estas_{\fdr_\numknockoffs}}
\newcommand{\newestas}{\estas_{\fdr,\operatorname{AKO}}}

\newcommand{\fdr}{q}

\Title{Aggregating Knockoffs for False Discovery Rate Control with an Application to Gut Microbiome Data
}

\TitleCitation{Aggregating Knockoffs for False Discovery Rate Control with an Application to Gut Microbiome Data}

\Author{Fang Xie *\orcidA{} and Johannes Lederer}
\AuthorNames{Fang Xie and Johannes Lederer}

\AuthorCitation{Xie, F.; Lederer, J.}
\address[1]{%
Department of Mathematics, Ruhr-University Bochum, Universit\"{a}tsstraße 150, 44801 Bochum, Germany; johannes.lederer@rub.de}
\corres{\hangafter=1 \hangindent=1.05em \hspace{-0.82em}
Correspondence: fang.xie@rub.de} %Tel.: (optional; include country code; if there are multiple corresponding authors, add author initials) +xx-xxxx-xxx-xxxx (F.L.)}

\abstract{%A single paragraph of about 200 words maximum. For research articles, abstracts should give a pertinent overview of the work. We strongly encourage authors to use the following style of structured abstracts, but without headings: (1) Background: place the question addressed in a broad context and highlight the purpose of the study; (2) Methods: describe briefly the main methods or treatments applied; (3) Results: summarize the article's main findings; (4) Conclusion: indicate the main conclusions or interpretations. The abstract should be an objective representation of the article, it must not contain results which are not presented and substantiated in the main text and should not exaggerate the main conclusions.
Recent discoveries suggest that our gut microbiome plays an important role in our health and wellbeing.
However, the gut microbiome data are intricate; 
for example, 
the microbial diversity in the gut makes the data high-dimensional. 
While there are dedicated high-dimensional methods, such as the lasso estimator,
they always come with the risk of false discoveries. 
Knockoffs are a recent approach to control the number of false discoveries.
In this paper, we show that knockoffs can be aggregated to increase power while retaining sharp control over the false discoveries.
We support our method both in theory and simulations,
and we show that it can lead to new discoveries on microbiome data from the American Gut Project.
In particular, our results indicate that several phyla that have been overlooked so far are associated with obesity.}


\begin{document}
%%%%%%%%%%%%%%%%%%%%%%%%%%%%%%%%%%%%%%%%%%
%-------------------------
%------Introduction------
%-------------------------
\section{Introduction}
Research on associations between the microbiome in the human gut and health and disease has surged in recent years~\citep{Evans2013TheGut,Huttenhower2012Structure,Koliada2017Association,Ley2006Microbial}. 
Data on the microbiome are abundant in view of citizen science endeavors such as the American Gut Project~\citep{Americangut},
but these non-standard ways of data collection limit data quality.

Another difficulty in the analysis of microbiome data is the high dimensionality,
which means that the number of parameters is large.
%this includes, in particular, settings where the number of parameters is of the same order or even exceeds the number of samples. 
The high-dimensionality is due to the diversity of the microbiome: 
at the phylum level,
there are typically several dozen types of microbes;
at the genus level,
there are even several hundred types of microbes.
Finding the microbes that are connected to a trait, therefore, requires the use of variable selection techniques.
Consequently, in view of the low data quality and the high dimensionality, research on microbiome data is in particular need for  controlling false discovery rates.

As a specific example where false discovery control can be important,
we are interested in finding phyla and genera of microbes that are related to obesity.
Successful detection of such groups of microbes could eventually lead to new means for weight control.
We model the task as a variable selection problem in logistic regression with the obesity as the dependent variable and the (log-transformed) %~\citep{Aitchison1982Statistical}) 
counts of the microbial relative abundancies as the variables.
The number of parameters, that is, the number of potential phyla/genera, is large,
while the number of actually relevant phyla/genera is assumed to be small;
hence, we speak of sparse, high-dimensional logistic regression.
A number of corresponding variable selection methods have been established,
including especially $\ell_1$-penalized logistic regression~\citep{Ng2004Feature}, %\jl{revised this: we can talk about it},
which can be equipped with knockoffs to do FDR %Define, if appropriate.
 control~\citep{Barber2015Controlling}.
However, in our application, 
these standard pipelines perform insufficiently:
for example, phyla that are known to be associated with obesity are missed,
and few phyla are selected in the first~place.

To increase the variable selection performance,
we propose a simple aggregation scheme. 
It consists of two steps:
the knockoff method is run $\numknockoffs$~times with a decreasing FDR target level,
and the selections are then combined.
We show that this aggregation scheme keeps FDR control intact while improving power in practice.

Our three key contributions are:
\begin{enumerate}
	\item We introduce an aggregation scheme that provably retains the original methods' guarantees---see Theorem~\ref{thm:ko+fdrguarantee}.
	\item We show numerically that the aggregation can increase the original methods' power---see  Sections~\ref{subsec:simlinear} and~\ref{subsec:simlogistic}.
	\item We show that the resulting pipelines for FDR control can be readily applied to empirical data and lead to new discoveries---see Section~\ref{subsec:realapplication}.
\end{enumerate}

The remainder of the paper is organized as follows.
In Section~\ref{sec:methodandtheory}, we introduce our pipeline and establish its theory. 
In Sections~\ref{subsec:simlinear} and~\ref{subsec:simlogistic}, we verify the accuracy of the pipeline in simulations.
In Section~\ref{subsec:realapplication}, we then show the usefulness of the pipeline in selecting phyla and genera connected with obesity.
In Section~\ref{sec:discussion}, we finally conclude with a brief discussion.

%------------------------------%
%---- Method and Theory-----%
%------------------------------%
\section{Methods and Theory}\label{sec:methodandtheory}
In this section, we introduce and study our aggregation scheme for knockoffs. 
We presume $\samplesize$ independent data pairs $(\designrow_1,\outcomes_1),\dots,(\designrow_\samplesize,\outcomes_\samplesize)$,
where each $\designrow_i\in\R^\numvar$ is a vector of variables and $\outcomes_i\in\R$ an outcome.
We keep the relationship between the outcomes and variables completely general at this point---
the relationship could be linear, logistic, or anything else---but we assume that this relationship is captured by a parameter~$\target\in\Rp$. 
Our target for inference is then the active set $\trueas:=\bslu j: \targets_j\neq0\bsru$.

Two important quality measures for an estimate $\estasq\bfl\design,\outcome\bfr$ of $\trueas$ are the FDR level
\begin{equation*}
\mathrm{FDR}:=\E\bflbbbb\frac{|\estasq\bfl\design,\outcome\bfr\backslash \trueas|}{|\estasq\bfl\design,\outcome\bfr|\vee 1}\bfrbbbb
\end{equation*}
and the power
\begin{equation*}
\mathrm{power}:=\E\bflbbbb\frac{|\estasq\bfl\design,\outcome\bfr\cap \trueas|}{|\estasq\bfl\design,\outcome\bfr|\vee 1}\bfrbbbb\,,
\end{equation*}
where $|\cdot|$ denotes the cardinality of a set.
Our focus is on estimators that provide FDR control at level $\fdr\in[0,1]$, that is,
\begin{equation}\label{ineq:KO-FDRcontrol}
\mathrm{FDR}\leq \fdr\,,
\end{equation}
while having large power.
In the following, we recall that the knockoff filter provides FDR control.
We then equip the knockoff filter with an aggregation step to improve its power.

%-------------------------------------%
%---A brief intro for knockoff filter---%
%-------------------------------------%
\subsection{A Brief Introduction to the Knockoff Filter}\label{subsec:knockofffilter}
The main ingredient of the knockoff method~\citep{Barber2015Controlling} is a  "knockoff version"~$\designknockoffs\in\Rnp$ of the design matrix~$\design$.
This new matrix~$\designknockoffs$ is essentially~$\design$ with additional noise such that (1)~the estimator can distinguish predictors from~$\design$ and~$\designknockoffs$ but (2)~both design matrices still have a similar correlation structure.
The idea is then to compare the selections of predictors in~$\designknockoffs$ and in~$\design$\ when estimating on the extended design~$[\designknockoffs\,\design]$. 

Denoting $\designknockoffs=(\designrowknockoffs_1,\ldots,\designrowknockoffs_\samplesize)\tp$ and $\design=(\designrow_1,\ldots,\designrow_\samplesize)\tp$,
the underlying assumption is that $\designrow_i\sim \mathcal{N}(\zero_\numvar, \CovMatrix)$ for a positive definite matrix~$\CovMatrix\in\Rpp$.
The knockoffs are then generated according to~\citep{Barber2018Robust}
\begin{equation}\label{designknockoffs}
\designrowknockoffs_i|\designrow_i\sim \mathcal{N}(\boldsymbol{\mu}_i, \boldsymbol{V})\ \ {\rm for\ each\ }i\in\bslu 1,\ldots,\samplesize\bsru
\end{equation}
with
\begin{align*}
\boldsymbol{\mu}_i&:=\designrow_i-\designrow_i\CovMatrix\inv\mathrm{diag}\{\diagv\}\,,\\
\boldsymbol{V}&:=2\mathrm{diag}\{\diagv\}-\mathrm{diag}\{\diagv\}\CovMatrix\inv\mathrm{diag}\{\diagv\}\,,
\end{align*}
where $\diagv\in\Rp$ is such that~$\boldsymbol{V}$ is positive definite.

The variable selection is then based on the estimator
\begin{equation} \label{def:penalizedestimator}
\estimator\bfl\penaltycoef,\design,\designknockoffs,\outcome\bfr\in\argmin\limits_{\regressor\in\R^{2\numvar}}\bslbbb\loglikfuc\bflbb\regressor|\bfl\design~\designknockoffs\bfr,\outcome\bfrbb+\sum_{j=1}^{2\numvar}\penaltyfun\bfl\regressors_j\bfr\bsrbbb\,,
\end{equation}
where $\loglikfuc:\Rp\mapsto\R$ is a loss function, $[\design~\designknockoffs]\in\R^{\samplesize\times 2\numvar}$ the extended design matrix,
and $\penaltyfun: \R\mapsto[0,\infty)$ a penalty function with tuning parameter $\penaltycoef>0$. 
Examples include the lasso (where $\loglikfuc\bflb\regressor|\bfl\design~\designknockoffs\bfr,\outcome\bfrb:=\normtwos{\outcome-\bfl\design~\designknockoffs\bfr\regressor}$ and $\penaltyfun\bfl\regressors_j\bfr:=\penaltycoef|\regressors_j|$) and the penalized logistic regression (where $\loglikfuc\bflb\regressor|\bfl\design~\designknockoffs\bfr,\outcome\bfrb=-\sum_{i =1}^{n}(\outcomes_i\designrow_i\tp\regressor-\ln(1+\exp(\designrow_i\tp\regressor)))$ and $\penaltyfun\bfl\regressors_j\bfr:=\penaltycoef|\regressors_j|$). 

The basic test statistics of the knockoff approach then capture the maximal tuning parameter of each variable entering the model:
\begin{align*}
\Zstatistics_{j}\bfl\design,\outcome\bfr&:=\sup\bslb\penaltycoef:\estimator_{j}\bfl\penaltycoef,\design,\designknockoffs,\outcome\bfr\neq0\bsrb\\
\Zknockoffs_{j}\bfl\design,\outcome\bfr&:=\sup\bslb\penaltycoef:\estimator_{p+j}\bfl\penaltycoef,\design,\designknockoffs,\outcome\bfr\neq0\bsrb
\end{align*}
for $j\in\bslu 1,\ldots,\numvar\bsru$. 
We surpress the dependence on $\bfl\design,\outcome\bfr$ in the following for notational ease. 
The final test statistic $\Wstatistics:=(\Wstatistics_{1},\ldots,\Wstatistics_\numvar)\tp$ then combines the basic test statistics into  
\begin{equation*}%\label{Wstatistics}
\Wstatistics_{j}:=\max\bslu \Zstatistics_{j}, \Zknockoffs_{j}\bsru\cdot\mathrm{sign}(\Zstatistics_{j}-\Zknockoffs_{j})\ \ \ {\rm for\ } j\in\bslu 1,\ldots,\numvar\bsru\,.
\end{equation*}

This statistic compares how much earlier the original predictor enters the model as compared to the fake predictor---or the other way around.
The threshold of the standard knockoff  procedure for a given FDR level $\fdr\in[0,1]$ is then
\begin{equation*}
\threshold_\fdr:=\min\bslbbb t\in\mathcal\Wstatistics: \frac{\#\bslu j:\Wstatistics_{j}\le-t\bsru}{\#\bslu j: \Wstatistics_{j}\ge t\bsru\vee 1}\le \fdr\bsrbbb\,,\\%\label{eq:threshold}\\
\end{equation*}
where $\mathcal\Wstatistics:=\bslu |\Wstatistics_{j}|: j\in\bslu 1,\ldots,\numvar\bsru\bsru$. 

The knockoff procedure has another variant, 
called knockoff+, 
which differs in the~threshold:
\begin{equation*}
\threshold_\fdr^+:=\min\bslbbb t\in\mathcal\Wstatistics: \frac{1+\#\bslu j:\Wstatistics_{j}\le-t\bsru}{\#\bslu j: \Wstatistics_{j}\ge t\bsru\vee 1}\le \fdr\bsrbbb\,.%\label{eq:threshold+}
\end{equation*}

These definitions finally yield the active sets
\begin{align*}
\estasq&:=\bslb j: \Wstatistics_{j}\ge\threshold_q\bsrb\,,\\%\label{eq:activeset}\\
\estasq^+&:=\bslb j: \Wstatistics_{j}\ge\threshold_q^+\bsrb\,.%\label{eq:activeset+}
\end{align*}

The active sets~$\estasq^+$ indeed provide FDR control at level~$\fdr$, that is, satisfy inequality~\eqref{ineq:KO-FDRcontrol}---see~\cite{Barber2015Controlling}, [Theorem~2];
the active sets~$\estasq$ provide an approximate version of it---see~\cite{Barber2015Controlling}, [Theorem~1]. 

%--------------------------------------%
%-------Aggregation Knockoffs-------%
%--------------------------------------%
\subsection{Aggregating Knockoffs}\label{subsec:aggregationknockoffs}
We now introduce the aggregation scheme and its theory.
The aggregation scheme applies the knockoff method $\numknockoffs$ times and combines the results: 

\textit{Step 1\label{step1}: Given a target FDR $\fdr\in[0,1]$, choose a sequence
	$\fdr_1,\ldots,\fdr_\numknockoffs\in[0,1]$~\label{choice-of-qi} such that $\fdr=\sum_{i=1}^\numknockoffs\fdr_i$.
	Apply the standard knockoff (or knockoff+) procedure $\numknockoffs$ times, once for each target FDR $\fdr_i$, 
	and denote the corresponding $\numknockoffs$-estimated active sets by $\estasqone,\ldots,\estasqk$ (or $\estasqone^+,\ldots,\estasqk^+$).}

\textit{Step 2: Combine these $\numknockoffs$-estimated active sets by taking the union:
	\begin{equation*}
	\newestas\bfl\numknockoffs\bfr:=\cup_{i =1}^{\numknockoffs}\estasqi\,
	\ \ \ ({\rm or}\ \newestas^+\bfl\numknockoffs\bfr:=\cup_{i =1}^{\numknockoffs}\estasqi^+)\,.
	\end{equation*}}
	
	The intuition behind this scheme is that increasing the number of knockoffs stabilizes the outcome and improves the power.
While applied here to the knockoff method, we emphasize that the aggregation scheme can be applied to every model and estimator as long as the there is a method for FDR control to start with. 
Hence, rather than the standard knockoffs as used below, we could equally well use 
model-X knockoffs~\citep{Candes2018Panning}, deep knockoffs~\citep{Romano2019Deep}, or KnockoffGAN~\citep{Jordon2019Knockoffgan}.

On the other hand,
the aggregation scheme retains the knockoffs' theoretical guarantees:
\begin{Theorem}\label{thm:ko+fdrguarantee}
	Given a target FDR level $\fdr\in[0,1]$, the set $\newestas^+\bfl\numknockoffs\bfr$ of the aggregation scheme for knockoff+ provides FDR control at level~$\fdr$:
	\begin{equation*}
	\E\bflbbbb\frac{|\newestas^+\bfl\numknockoffs\bfr\backslash \trueas|}{|\newestas^+\bfl\numknockoffs\bfr|\vee 1}\bfrbbbb\le q\,.
	\end{equation*}
\end{Theorem}

Similarly, the scheme retains the approximate FDR control of standard knockoffs---we will demonstrate this in our simulations.

%The exact and theoretical FDR guarantee for the standard knockoff is still an open question. 
%If this question is solved, the similar result in Theorem~\ref{thm:ko+fdrguarantee} for standard knockoff can be obtained similarly.

\begin{proof}[Proof of Theorem~\ref{thm:ko+fdrguarantee}]
	The proof is---maybe suprisingly---simple.
	By Theorem 2 in~\citep{Barber2015Controlling}, we have for all $\fdr_i$, $i\in\bslu 1,\ldots,\numknockoffs\bsru$,
	\begingroup
	\allowdisplaybreaks
	\begin{equation*}%\label{ineq:FDRcontrol-qi}
	\E\bflbbbb\frac{|\estasqi^+\backslash \trueas|}{|\estasqi^+|\vee 1}\bfrbbbb\le \fdr_i\,.
	\end{equation*}
	
	Hence,
	\begin{align*}
	\E\bflbbbb\frac{|\newestas^+\bfl\numknockoffs\bfr\backslash \trueas|}{|\newestas^+\bfl\numknockoffs\bfr|\vee 1}\bfrbbbb&=\E\bflbbbb\frac{|(\cup_{i =1}^{\numknockoffs}\estasqi^+)\backslash \trueas|}{|\cup_{i =1}^{\numknockoffs}\estasqi^+|\vee 1}\bfrbbbb\\
	&=\E\bflbbbb\frac{|\cup_{i =1}^{\numknockoffs}(\estasqi^+\backslash \trueas)|}{|\cup_{i =1}^{\numknockoffs}\estasqi^+|\vee 1}\bfrbbbb\\
	&\le\E\bflbbbb\sum_{i =1}^{\numknockoffs}\frac{|\estasqi^+\backslash \trueas|}{|\cup_{i =1}^{\numknockoffs}\estasqi^+|\vee 1}\bfrbbbb\\
	&\le\E\bflbbbb\sum_{i =1}^{\numknockoffs}\frac{|\estasqi^+\backslash \trueas|}{|\estasqi^+|\vee 1}\bfrbbbb\\
	&=\sum_{i =1}^{\numknockoffs}\E\bflbbbb\frac{|\estasqi^+\backslash \trueas|}{|\estasqi^+|\vee 1}\bfrbbbb\\
	&\le\sum_{i =1}^{\numknockoffs} \fdr_i
	=\fdr\,,
	\end{align*}
	%	\begin{align*}
	%	\E\bflbbbb\frac{|\newestas^+\bfl\numknockoffs\bfr\backslash \trueas|}{|\newestas^+\bfl\numknockoffs\bfr|\vee 1}\bfrbbbb&=\E\bflbbbb\frac{|(\cup_{i =1}^{\numknockoffs}\estasqi^+)\backslash \trueas|}{|\cup_{i =1}^{\numknockoffs}\estasqi^+|\vee 1}\bfrbbbb\\
	%	&=\E\bflbbbb\frac{|\cup_{i =1}^{\numknockoffs}(\estasqi^+\backslash \trueas)|}{|\cup_{i =1}^{\numknockoffs}\estasqi^+|\vee 1}\bfrbbbb\\
	%	&\le\E\bflbbbb\sum_{i =1}^{\numknockoffs}\frac{|\estasqi^+\backslash \trueas|}{|\cup_{i =1}^{\numknockoffs}\estasqi^+|\vee 1}\bfrbbbb\\
	%	&\le\E\bflbbbb\sum_{i =1}^{\numknockoffs}\frac{|\estasqi^+\backslash \trueas|}{|\estasqi^+|\vee 1}\bfrbbbb\\
	%	&=\sum_{i =1}^{\numknockoffs}\E\bflbbbb\frac{|\estasqi^+\backslash \trueas|}{|\estasqi^+|\vee 1}\bfrbbbb\\
	%	&\le\sum_{i =1}^{\numknockoffs} \fdr_i\\
	%	&=\fdr\,,
	%	\end{align*}
	\endgroup
	as desired.
\end{proof}

For $\numknockoffs=1$, 
our method equals the standard knockoff (or knockoff+) procedure;
in practice, we recommend $\numknockoffs\approx5$ as a trade-off between computational effort and statistical effect. 
We also recommend $\fdr_i=\fdr/2^{i-1}$,
which, strictly speaking, does not meet our assumptions on $\fdr_1,\ldots,\fdr_\numknockoffs$ exactly, but it works well empirically---see the simulations.

\subsection{Other Approaches}
While preparing this manuscript, 
two other ways of aggregating knockoffs were proposed~\citep{Holden2018Multiple,Gimenez2019Improving}. 
While we use $k$ knockoffs in $k$ processes individually and aggregate  at the end,
they use multiple knockoffs simultaneously in one process.
They can also show that their schemes provide valid FDR control,
but we can argue that our approach is considerably simpler.
We illustrate in the Appendix~\ref{Appendix:subsec:MKO} that we typically get more power than the multiple knockoffs method proposed by~\citep{Holden2018Multiple};
we have not gotten the scheme of~\citep{Gimenez2019Improving} to run yet.

%--------------------------%
%-------Simulations-------%
%--------------------------%
\section{Simulations and a Real Data Analysis}\label{sec:simulationsandrealdata}
In this section,
we test our method empirically both on synthetic and on real data.
We incorporate the specifics of microbiome data: 
First, in line with recent proposals for generalized linear models in this context~\citep{Lu2019Generalized},
we study linear regression as well as logistic regression,
and we transform count data with the standard log-transformation~\citep{Aitchison1982Statistical}. 
Second, since gut microbiome data tends to be correlated~\citep{Naqvi2010Network},
we ensure that the synthetic data are also correlated.
Third, since gut microbime data also tend to be zero-inflated,
we replace zero values by 0.5 times the observed minimum abundance,
which is standard approach in microbiome analysis~\citep{Aitchison2003Statistical, Kurtz2015Sparse}. 
We compare our modified aggregating knockoff pipeline applied to $\ell_1$-penalized regression (called AKO henceforth) with the standard knockoff pipeline (KO). 
%In addition, we also add comparison to multiple knockoffs method (MKO) of~\citep{Holden2018Multiple} for two simulations.

Throughout, we set $\numknockoffs=5$ and $\fdr_i=\fdr/2^{i-1}$. 
%To be fair, we also set $k=5$ for the MKO. 
We also show the results for other choice of $\fdr_1,\dots,\fdr_k$ in the Appendix~\ref{Appendix:subsec:choice-of-qi}.

\subsection{Simulation 1: Linear Regression}\label{subsec:simlinear}
We first consider linear regression.
The dimensions of the synthetic data are $(\samplesize,\numvar)\in\bslu(200,100),(400,200)\bsru$. 
The rows~$\designrow_i$ of the design matrix  are sampled i.i.d. %Define, if appropriate.
 from $\mathcal{N}(\zero_\numvar, \boldsymbol{\Sigma})$ with the elements in $\boldsymbol{\Sigma}$ satisfying $\Sigma_{ij}=\rho^{|i-j|}$ with correlation factor $\rho=0.5$. 
The noise is drawn from $\noise\sim\mathcal{N}(\zero, \variance\identitymatrix_\samplesize)$. 
The true parameter $\target$ has $20$ nonzero coefficients, which are selected uniformly at random from $\bslu1,\ldots,\numvar\bsru$ and set to~$1$  before the entire vector is rescaled to have signal-to-noise ratio ${\normtwos{\design\target}}/n\variance=5$ with $\variance=1$. 
The outcome $\outcome$ is then generated by the linear model
\begin{equation*}
\outcome = \design\target+\noise.
\end{equation*}

The test statistic $\Wstatistics$ is based on the lasso method as described in the preceeding section.

We compare the empirical FDR and power averaged over $\numFDR=100$ repetitions of the simulations for each method.
%\begin{align*}
% \langle \operatorname{FDR} \rangle_{\mathrm{AKO}}&:=\frac1\numFDR\sum_{m=1}^{\numFDR}\frac{|\newestas^m\bfl\numknockoffs\bfr\backslash \trueas|}{|\newestas^m\bfl\numknockoffs\bfr|\vee 1}\\
% \langle \operatorname{power} \rangle&:=\frac{1}{\numFDR}\sum_{m=1}^\numFDR\frac{|\newestas^m\bfl\numknockoffs\bfr\cap \trueas|}{|\trueas|\vee 1}\,
% \end{align*}
The ideal result would be an average FDR of at most~$\fdr$ and a power equal to 1.

\begin{figure}[H]
%	\centering
	\subfigure[$(\samplesize,\numvar)=(200,100)$]{
		\label{fig:n=200}
		\hfill
		\begin{minipage}{0.24\textwidth}
			\includegraphics[width=\textwidth]{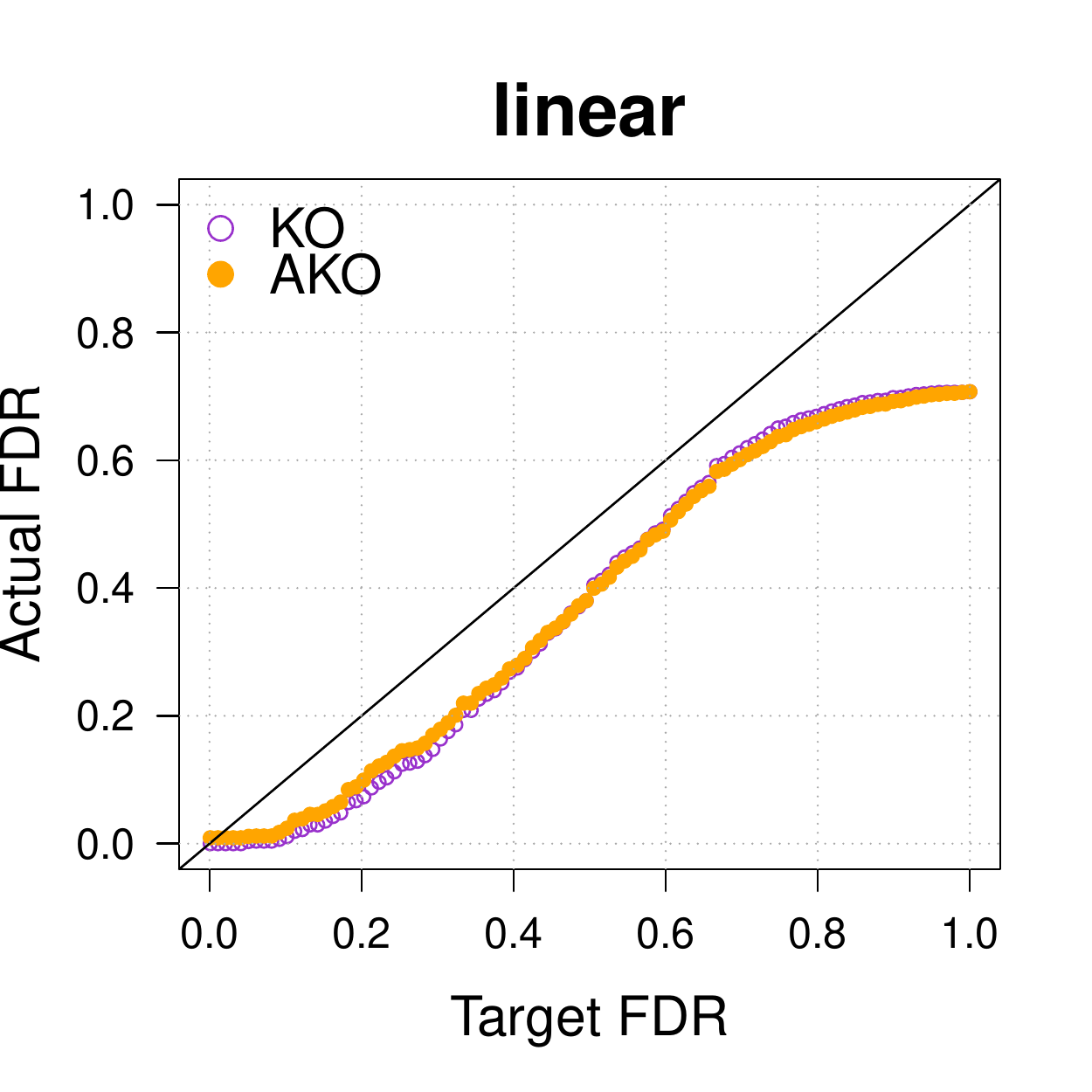}
		\end{minipage}
		\begin{minipage}{0.24\textwidth}
			\includegraphics[width=\textwidth]{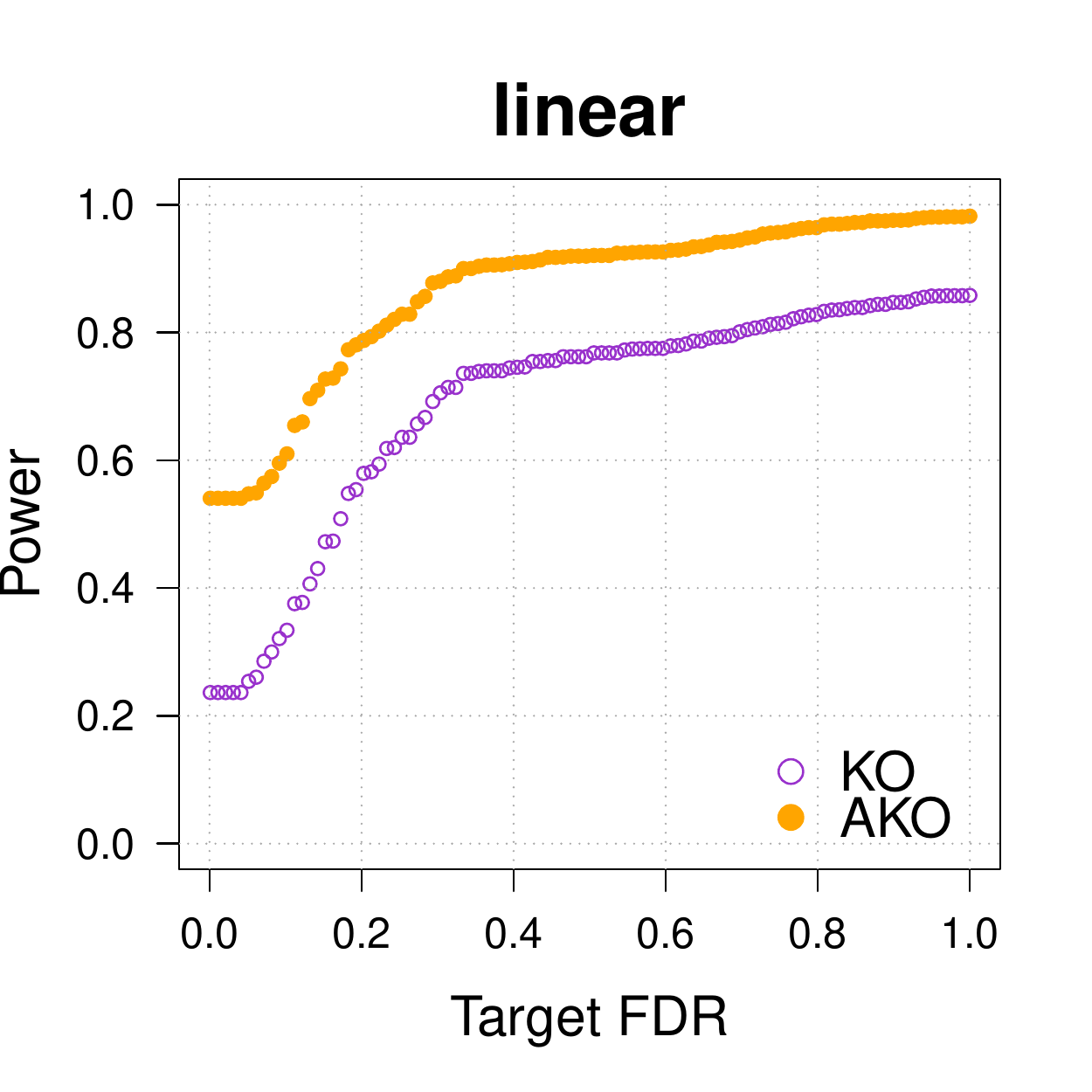}
	\end{minipage}}
	\subfigure[$(\samplesize,\numvar)=(400,200)$]{
		\label{fig:n=400}
		\begin{minipage}{0.24\textwidth}
			\includegraphics[width=\textwidth]{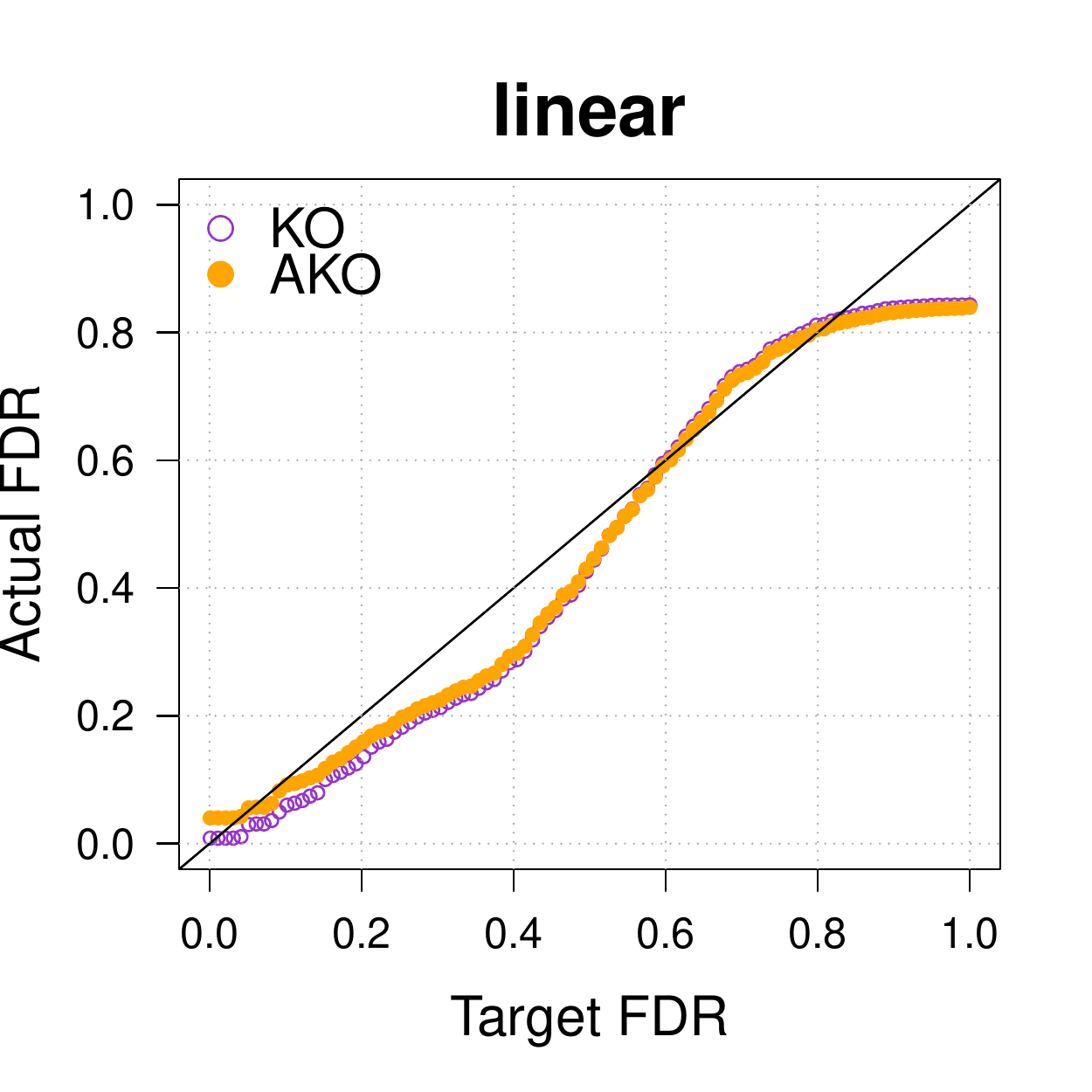}
		\end{minipage}
		\begin{minipage}{0.24\textwidth}
			\includegraphics[width=\textwidth]{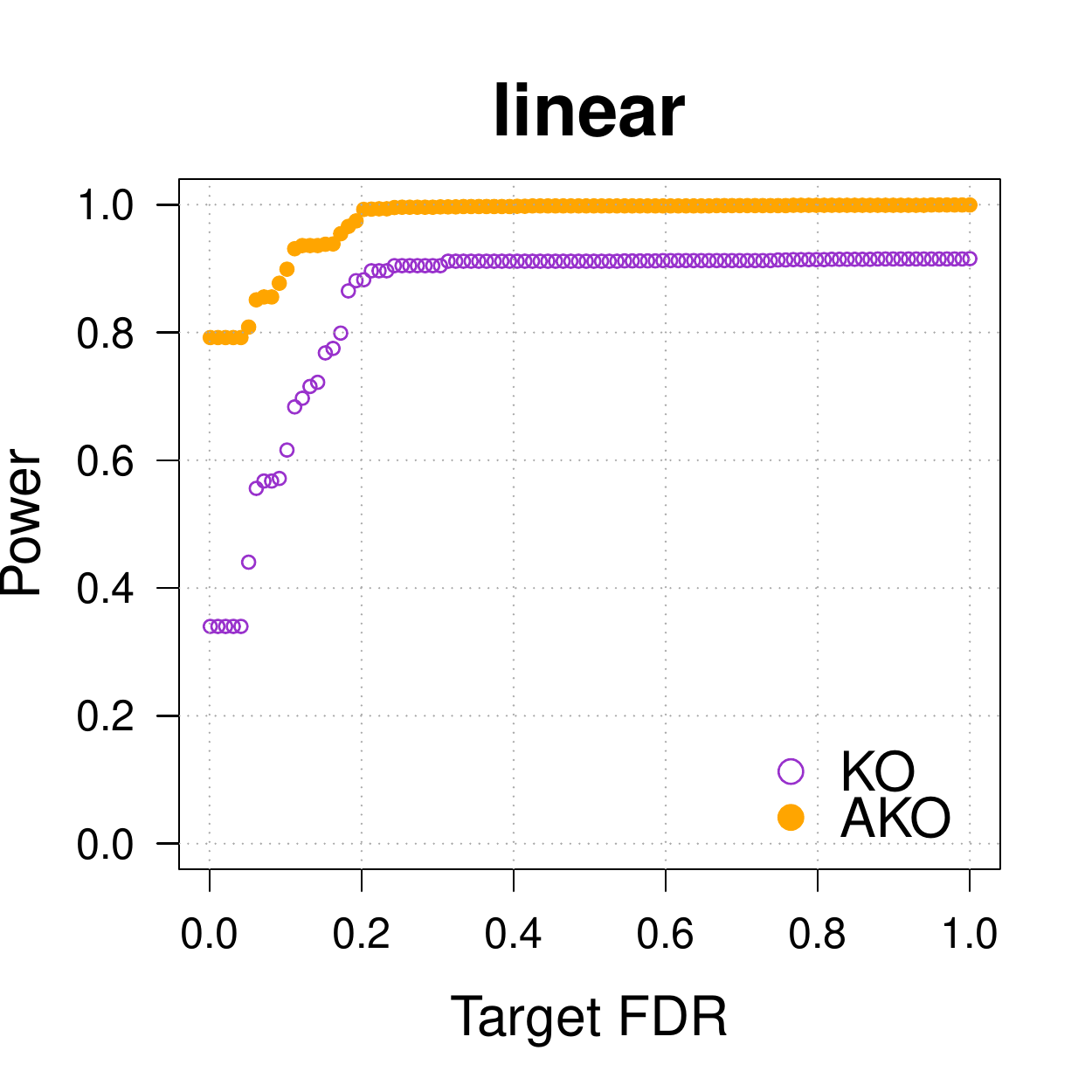}
	\end{minipage}}
	\caption{Our approach, AKO %Define, if appropriate.
		(solid, orange circles), has a similar FDR %Define, if appropriate.
		to the standard KO (hollow, purple circles) but has more power.}%
	\label{fig1}%
\end{figure}

The results in Figure~\ref{fig1}a,b show that our pipeline retains the KO's FDR control while increasing the power. 
%The MKO is more conservative than both our pipeline and the KO, and our pipeline has higher power comparing to the MKO when the target FDR is small.

% start a new page without indent 4.6cm
%\clearpage
\end{paracol}
\nointerlineskip

\begin{figure}[H]
\widefigure
%	\centering
	\subfigure[$(\samplesize,\numvar)=(200,100)$]{
		\label{fig:logisticn=200}
		\begin{minipage}{0.24\textwidth}
			\includegraphics[width=\textwidth]{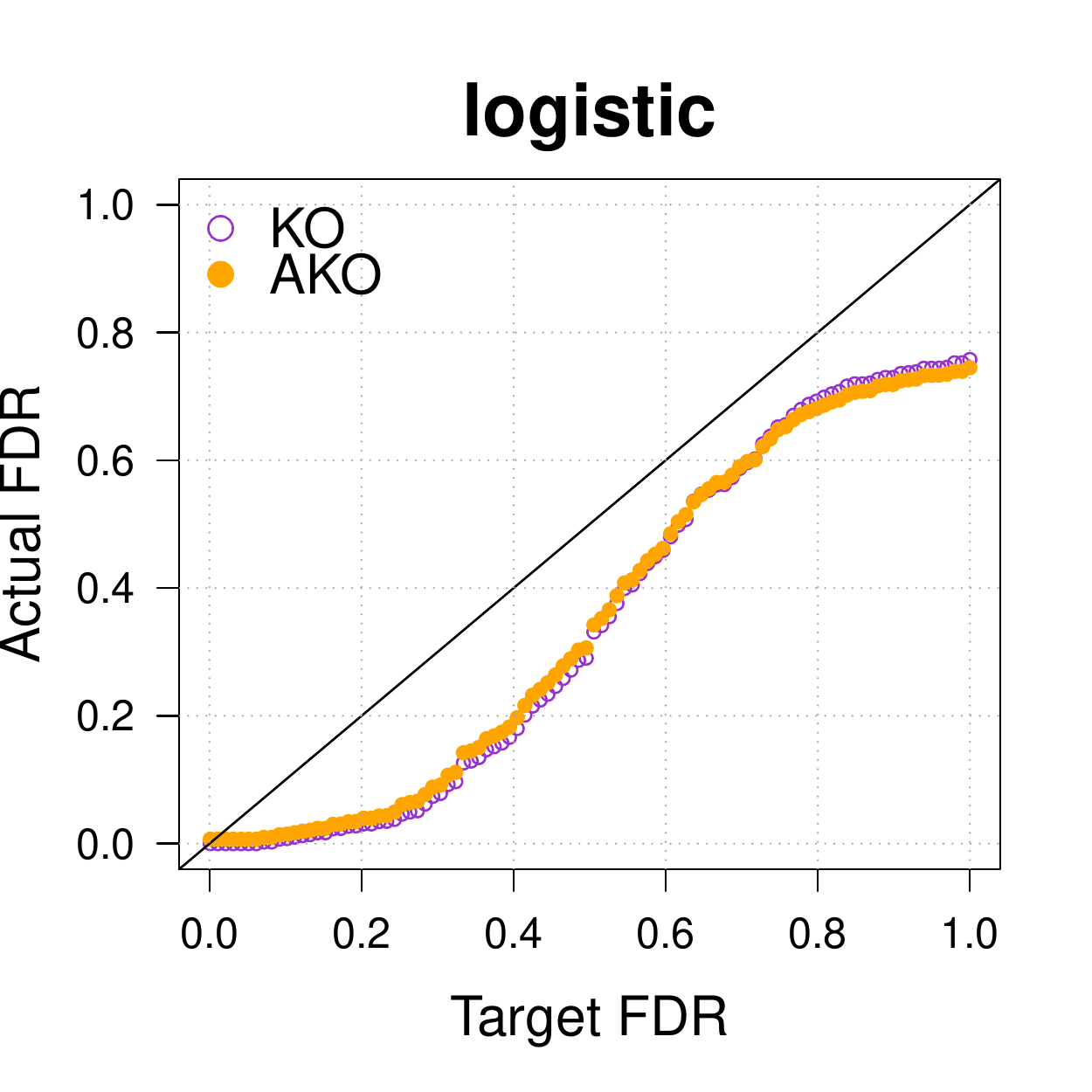}
		\end{minipage}
		\begin{minipage}{0.24\textwidth}
			\includegraphics[width=\textwidth]{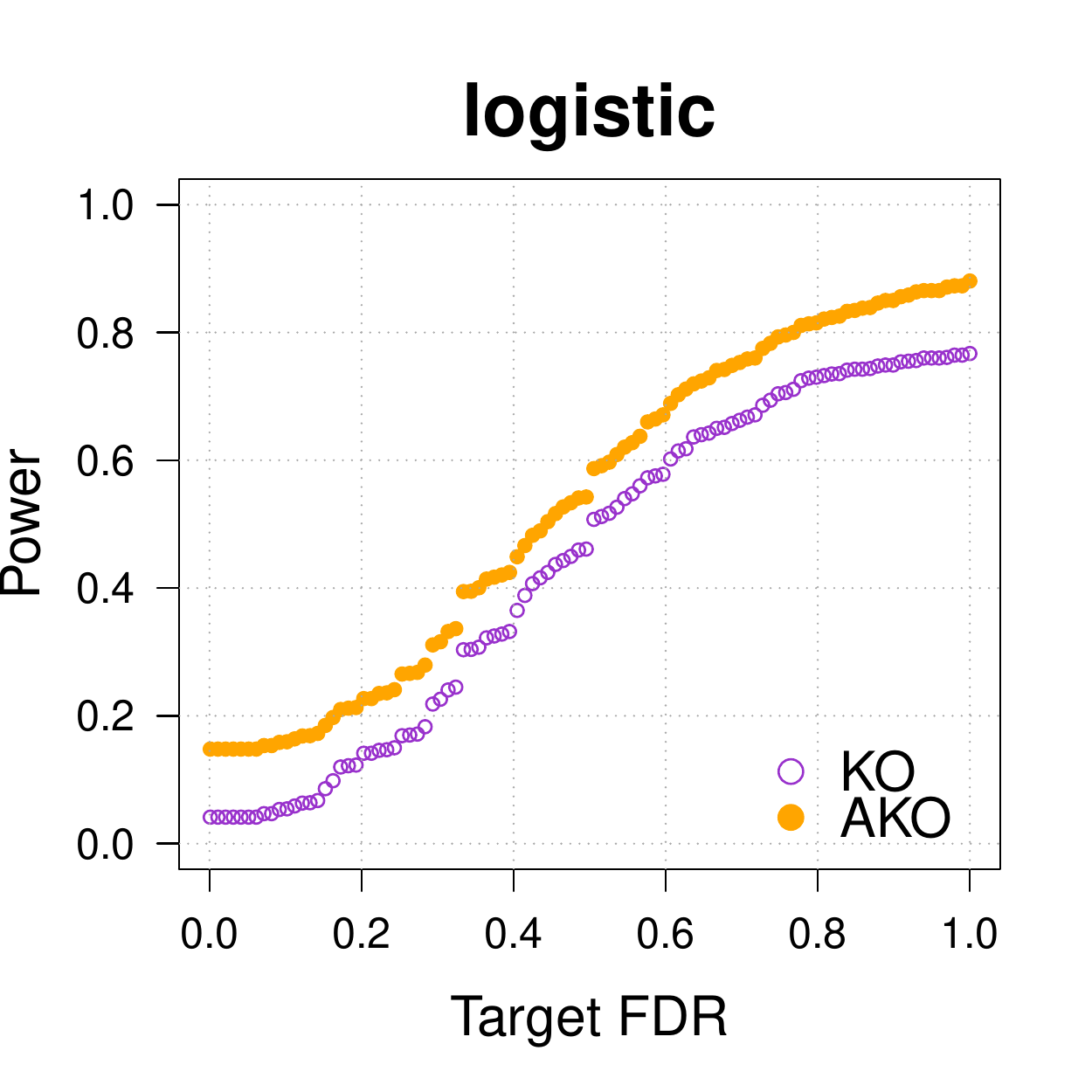}
		\end{minipage}
	}
	\subfigure[$(\samplesize,\numvar)=(400,200)$]{
		\label{fig:logisticn=400}
		\begin{minipage}{0.24\textwidth}
			\includegraphics[width=\textwidth]{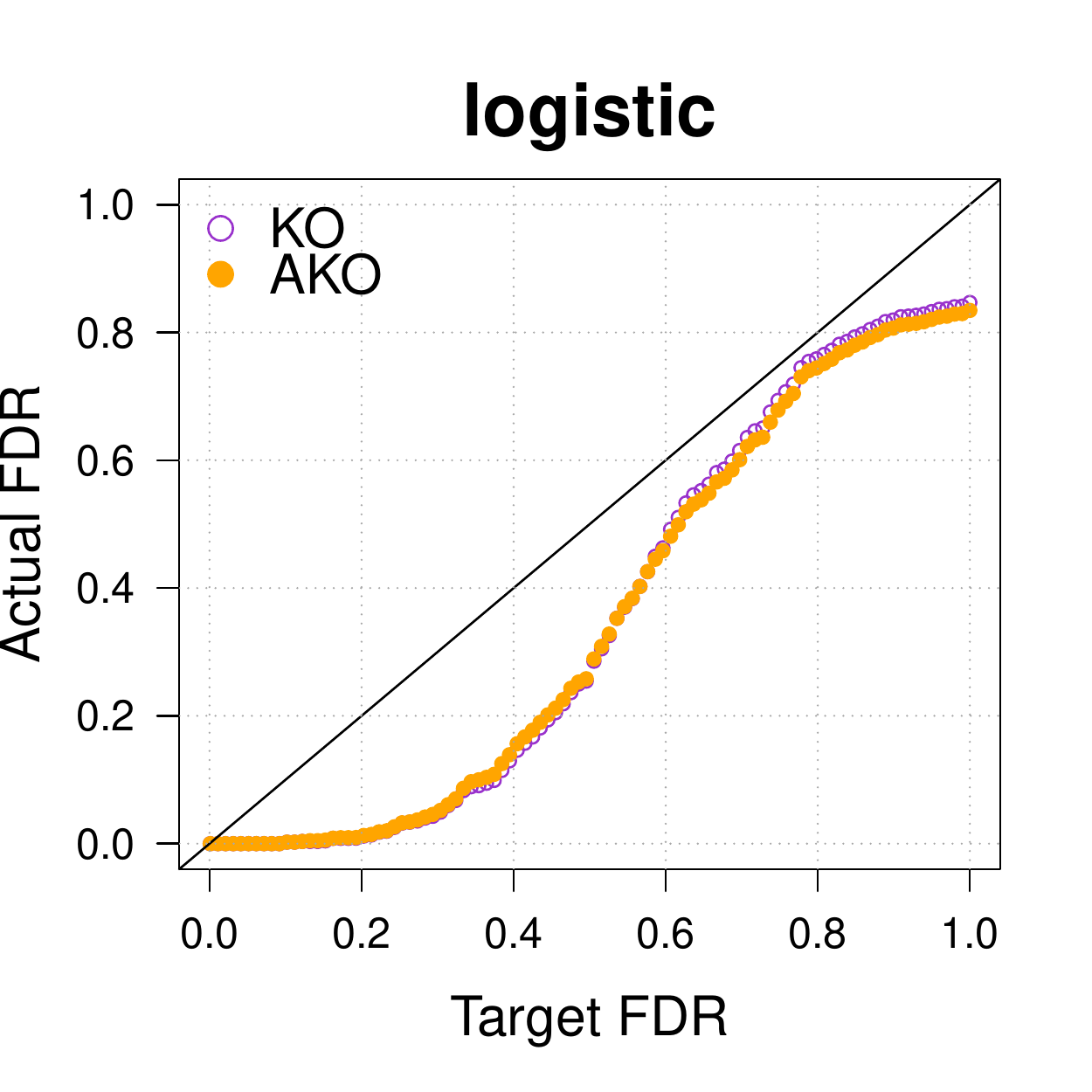}
		\end{minipage}
		\begin{minipage}{0.24\textwidth}
			\includegraphics[width=\textwidth]{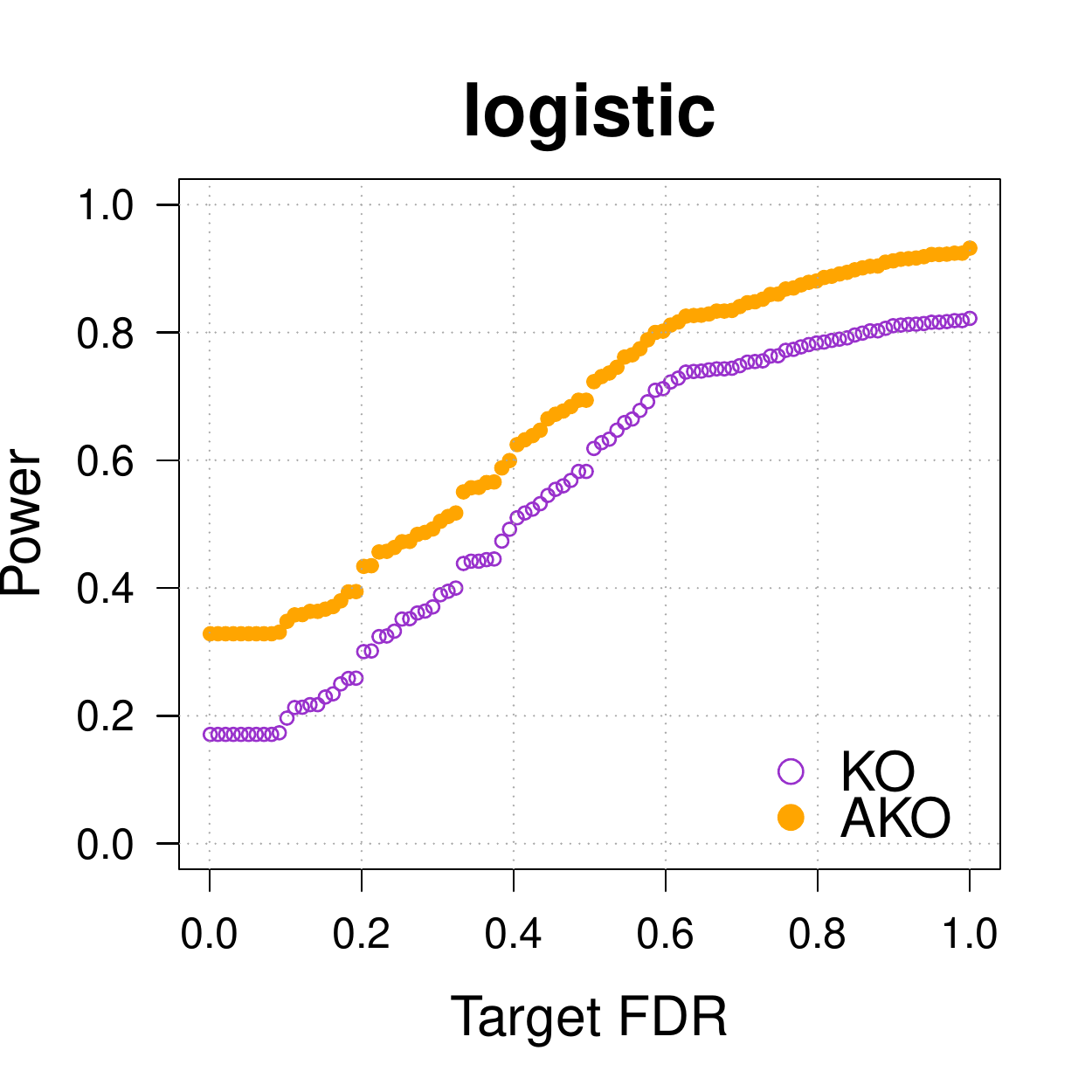}
		\end{minipage}
	}
	\caption{Our approach AKO (solid, orange circles) has a similar FDR to the standard KO (hollow, purple circles) but has more power. \label{fig2}}%
\end{figure}

\begin{paracol}{2}
%\linenumbers
\switchcolumn

\subsection{Simulation 2: Logistic Regression}\label{subsec:simlogistic}
We now consider logistic regression (the paper~\citep{Klose2020Pipeline} was the first one to apply our scheme to logistic regression).
The above settings remain the same except for the outcomes $\outcomes_i$ being generated by
\begin{equation*}
\Pr(\outcomes_i=1)=\frac{\exp(\designrow_i\tp\target)}{1+\exp(\designrow_i\tp\target)}\ \ \ {\rm for}\ i\in\bslu 1,\ldots,\samplesize\bsru.
\end{equation*}

The results in Figure~\ref{fig2}a,b show again that our pipeline retains the KO's FDR control while increasing the power. 
%The MKO is more conservative than both our pipeline and the KO, and our pipeline has higher power comparing to the MKO. 
For more simulations in various settings, please refer to Appendix~\ref{Appendix:subsec:various-settings}.

% start a new page without indent 4.6cm
%\clearpage
\end{paracol}
\nointerlineskip

\begin{paracol}{2}
%\linenumbers
\switchcolumn

%--------------------------%
%----Real Application------%
%--------------------------%

\subsection{Influence of the Gut Microbiome on Obesity}\label{subsec:realapplication}
A well-functioning gut microbiome is essential for health~\citep{Huttenhower2012Structure};
for example, there is strong evidence that the composition of the microbiome is connected to obesity~\citep{Escobar2014The,Gerard2016Gut,Turnbaugh2009The}. 
Existing research about this connection has focused only on phyla of bacteria that are abundant in most guts, such as \emph{Actinobacteria}, \emph{Bacteroidetes}, \emph{Firmicutes}, and \emph{Verrucomicrobia}~\citep{Bai2019Composition,Clarke2012Gut,Depommier2019Supplementation,Gao2018Body,Koliada2017Association}.
In the following analysis, in contrast, we include the microbiome in its entirety.
Our findings suggest that also phyla beyond those ones mentioned in the literature are connected to obesity.

The data for our analysis are from the American Gut Project~\citep{Americangut}. 
The scope is American adults with age between~20 and 69 and BMI between $15\,\mathrm{kg/m^2}$ and $60\,\mathrm{kg/m^2}$ in the collection up to January, 2018.
This includes  $n=8404$ subjects, of which
$278$~are underweight (uw; BMI below $18.5\,\mathrm{kg/m^2}$), 
$4972$~have normal weight (nor; 18.5--25$\,\mathrm{kg/m^2}$), 
$2253$~are overweight (ow; 25--30$\,\mathrm{kg/m^2}$), 
and $901$~are obese (ob; above $30\,\mathrm{kg/m^2}$). 
We transform the data and deal with the zero-counts as decribed earlier. 
The total number of phyla in our scope is~$p=55$.

The underlying model for these data is $\ell_1$-penalized logistic regression model with outcomes $\outcomes_i=1$ (ob) when BMI $\ge30$ and $\outcomes_i=0$ (non-ob) otherwise. 
The target FDR level is~$\fdr=0.1$.
Seven different groupings are considered to get the most out of the data: 
(i) all four groups (uw+nor+ow+ob), (ii) uw+ob, (iii) nor+ob, (iv) ow+ob, (v) uw+nor+ob, (vi)~uw+ow+ob and (vii) nor+ow+ob. 
The AKO is applied with $\bslu\fdr_i=\fdr/2^{i-1}: i\in\bslu 1,\ldots,k\bsru\bsru$.

The results in Table~\ref{tab:realdatalogistic} show that our pipeline selects more phyla in general. 
Since the theory and the above similations suggest that both methods have similar FDR,
the results indicate that our pipeline has more power.
In particilar, phyla that are known to be connected with obesity, such as \emph{Actinobacteria}, \emph{Bacteroidetes}, \emph{Firmicutes}, and \emph{Verrucomicrobia}~\citep{Bai2019Composition,Clarke2012Gut,Depommier2019Supplementation,Gao2018Body,Koliada2017Association}, are selected by AKO more often across the seven groupings.

The results also highlight \emph{Spirochaetes},
which have not been associated with obesity in the literature, yet.
The standard pipeline does not seem to have the power to select it,
and similarly, 
two other known FDR control methods for microbiome data---the standard BH procedure~\citep{BH95} and the TreeFDR~\citep{Xiao2017False}---select \emph{Spirochaetes} only for large FDR levels ($q\gtrsim 0.8$).
(Futher comparisons with the BH procedure, TreeFDR, and the plain KO can be found in the Appendix~\ref{Appendix:subsec:competitors}.)
In contrast, our method selects \emph{Spirochaetes} even at very small FDR levels (such as $q=0.01$),
which strongly suggests a connection between  \emph{Spirochaetes}  and obesity.

Our pipeline can, of course, also be applied to more detailed taxonomic ranks.
As an illustration, we report the results of an application to genera for ALL---cf.\@ (i) in Table~\ref{tab:realdatalogistic}---in Table~\ref{tab:genera}. The data sampling is same with the phyla analysis. 
The only difference is that the total number of genera is $\numvar = 969$. 
The rest of results for other six different groupings are given in the Appendix~\ref{Appendix:genera}. 
We find correspondingly that AKO selects more genera than the standard KO.

% start a new page without indent 4.6cm
%\clearpage
\end{paracol}
\nointerlineskip
\begin{specialtable}[H]
\tablesize{\small}	
\widetable	%MDPI: Is the italics necessary? if not, we will remove %FX and JL: You can remove it.
	\caption{{Selected bacterial phyla by our pipeline} (AKO) and the original pipeline (KO) at FDR level $\fdr=0.1$ for seven groupings.
		AKO consistently selects more phyla than KO.} \label{tab:realdatalogistic}
	\centering
%	\tiny
	\setlength{\cellWidtha}{\columnwidth/4-2\tabcolsep+0in}
\setlength{\cellWidthb}{\columnwidth/4-2\tabcolsep+0in}
\setlength{\cellWidthc}{\columnwidth/4-2\tabcolsep-0in}
\setlength{\cellWidthd}{\columnwidth/4-2\tabcolsep-0in}
\scalebox{1}[1]{\begin{tabularx}{\columnwidth}{>{\PreserveBackslash\centering}p{\cellWidtha}>{\PreserveBackslash\centering}p{\cellWidthb}>{\PreserveBackslash\centering}p{\cellWidthc}>{\PreserveBackslash\centering}p{\cellWidthd}}
\toprule
		\multicolumn{2}{c}{\textbf{(i) all}} & \multicolumn{2}{c}{\textbf{(ii) uw + ob}}\\
		\textbf{KO} &  \textbf{AKO} & \textbf{KO} &  \textbf{AKO}\\
		\hline
		\emph{Actinobacteria} & \emph{Actinobacteria} & \emph{Actinobacteria} & \emph{Actinobacteria}\\
		& \emph{Bacteroidetes} & &   \\ 
		\emph{Cyanobacteria} & \emph{Cyanobacteria} &  &  \emph{Cyanobacteria} \\
		&  &   & \emph{Firmicutes} \\
		\emph{Proteobacteria} & \emph{Proteobacteria} & &  \\ 
		&  \emph{Spirochaetes}& &  \\ 
		\emph{Synergistetes} & \emph{Synergistetes} &  & \emph{Synergistetes} \\
		\emph{Tenericutes} & \emph{Tenericutes} & \emph{Tenericutes}  & \emph{Tenericutes} \\
		& \emph{Verrucomicrobia} &  &  \\ 
		\hline
		\multicolumn{2}{c}{\textbf{(iii) nor + ob}} & \multicolumn{2}{c}{\textbf{(iv) ow + ob}}\\
		\textbf{KO}  & \textbf{AKO} & \textbf{KO}  & \textbf{AKO} \\
		\hline
		\emph{Actinobacteria} & \emph{Actinobacteria} & & \emph{Actinobacteria} \\ 
		\emph{Bacteroidetes} & \emph{Bacteroidetes} & &  \\ 
		\emph{Cyanobacteria} & \emph{Cyanobacteria} & \emph{Cyanobacteria} & \emph{Cyanobacteria} \\
		&  &  & \emph{Firmicutes} \\
		& \emph{Lentisphaerae} &  & \\
		\emph{Proteobacteria} & \emph{Proteobacteria} & & \emph{Proteobacteria} \\ 
		&  \emph{Spirochaetes} & &  \emph{Spirochaetes} \\ 
		\emph{Synergistetes} & \emph{Synergistetes} &   & \emph{Synergistetes} \\
		&  &  & \emph{TM7} \\ 
		\emph{Tenericutes} & \emph{Tenericutes} & \emph{Tenericutes} & \emph{Tenericutes} \\
		&  \emph{Verrucomicrobia} & & \\ 
		&  \emph{Thermi} & &  \\ 
		\hline
		\multicolumn{2}{c}{\textbf{(v) uw + nor + ob}} & \multicolumn{2}{c}{\textbf{(vi) uw + ow + ob}}\\
		\textbf{KO} & \textbf{AKO} & \textbf{KO} & \textbf{AKO} \\
		\hline 
		\emph{Actinobacteria} & \emph{Actinobacteria} & &\emph{Actinobacteria} \\
		\emph{Bacteroidetes}  & \emph{Bacteroidetes} & &   \\ 
		\emph{Cyanobacteria} & \emph{Cyanobacteria} & \emph{Cyanobacteria} & \emph{Cyanobacteria} \\
		&   &  & \emph{Firmicutes} \\
		& \emph{Lentisphaerae} &  & \\
		\emph{Proteobacteria} & \emph{Proteobacteria} & & \emph{Proteobacteria} \\
		& \emph{Spirochaetes} & & \emph{Spirochaetes} \\  
		\emph{Synergistetes} & \emph{Synergistetes} &  & \emph{Synergistetes} \\
		&   &  & \emph{TM7} \\ 
		\emph{Tenericutes} & \emph{Tenericutes} & \emph{Tenericutes} & \emph{Tenericutes} \\
		\hline
		\multicolumn{2}{c}{\textbf{(vii) nor+ow+ob}} & \multicolumn{2}{c}{}\\
		\textbf{KO} & \textbf{AKO} &  & \\
		\hline
		& \emph{Actinobacteria}  &   &    \\
		&  \emph{Bacteroidetes} & &  \\ 
		\emph{Cyanobacteria} & \emph{Cyanobacteria} & & \\ 
		\emph{Proteobacteria} & \emph{Proteobacteria} & & \\ 
		& \emph{Spirochaetes} &  &   \\
		\emph{Synergistetes} & \emph{Synergistetes} & &  \\ 
		\emph{Tenericutes} & \emph{Tenericutes} & & \\ 
		& \emph{Verrucomicrobia} & & \\ 
		\bottomrule
	\end{tabularx}}
\end{specialtable}
\begin{paracol}{2}
%\linenumbers
\switchcolumn

% genus level
\begin{specialtable}[H]
\tablesize{\small}	 %MDPI: Is the italics necessary? if not, we will remove
%\widetable	  
	\caption{{Selected bacterial genera by our pipeline} (AKO) and the original pipeline (KO) at FDR level $\fdr = 0.1$ for ALL---cf.\@ (i) in Table~\ref{tab:realdatalogistic}.
		AKO selects more genera than the original KO.} \label{tab:genera}
%	\centering
%	\tiny
	\setlength{\cellWidtha}{\columnwidth/3-2\tabcolsep+0in}
\setlength{\cellWidthb}{\columnwidth/3-2\tabcolsep+0in}
\setlength{\cellWidthc}{\columnwidth/3-2\tabcolsep-0in}
\scalebox{1}[1]{\begin{tabularx}{\columnwidth}{>{\PreserveBackslash\centering}p{\cellWidtha}>{\PreserveBackslash\centering}p{\cellWidthb}>{\PreserveBackslash\centering}p{\cellWidthc}}
\toprule

		%Phylum & \multicolumn{3}{c}{Genus}\\
		%\hline
		\textbf{Phylum} & \textbf{KO} & \textbf{AKO}\\
		\hline 
		\multirow{3}{*}{\emph{Actinobacteria}} & \emph{Actinomyces} & \emph{Actinomyces}\\
		& \emph{Collinsella} & \emph{Collinsella} \\ 
		& \emph{Eggerthella} & \emph{Eggerthella} \\
		\hline  
		\multirow{2}{*}{\emph{Cyanobacteria}} & \emph{YS2} & \emph{YS2} \\ 
		& & \emph{Streptophyta} \\
		\hline
		\multirow{12}{*}{\emph{Firmicutes}} & \emph{Bacillus} & \emph{Bacillus} \\
		& & \emph{Lactobacillus} \\ 
		&  \emph{Lactococcus} & \emph{Lactococcus} \\
		& & \emph{Clostridium} \\
		& \emph{Lachnospira} & \emph{Lachnospira} \\
		& \emph{Ruminococcus} & \emph{Ruminococcus} \\ 
		& & \emph{Peptostreptococcaceae} \\
		& \emph{Acidaminococcus} & \emph{Acidaminococcus} \\
		& \emph{Megasphaera} & \emph{Megasphaera} \\
		& & \emph{Mogibacteriaceae} \\
		& \emph{Erysipelotrichaceae} & \emph{Erysipelotrichaceae} \\ 
		& \emph{Catenibacterium} & \emph{Catenibacterium}\\
		\hline
		\multirow{2}{*}{\emph{Proteobacteria}} & \emph{RF32} & \emph{RF32} \\ 
		& \emph{Haemophilus} & \emph{Haemophilus} \\
		\hline
		\emph{Tenericutes} & & \emph{RF39} \\
		\bottomrule
	\end{tabularx}}
\end{specialtable}

%--------------------------%
%--------Conclusion-------%
%--------------------------%

\section{Discussion}\label{sec:discussion}
Our aggregation scheme for knockoffs is supported by theory (Section~\ref{sec:methodandtheory}) and simulations (Sections~\ref{subsec:simlinear} and~\ref{subsec:simlogistic})
and may lead to new discoveries in microbiomics \mbox{(Section~\ref{subsec:realapplication}).}

While we focus on a specific pipeline,
our concept applies very generally.
For example, it does not depend on the underlying statistical model or estimator but only on the availability of FDR control.
In particular, the FDR control can be established via standard knockoffs or any other scheme.
This flexibility is particularly interesting in practice:
while the standard knockoffs rely on normality,
other knockoff procedures, such as model-X knockoffs~\citep{Candes2018Panning}, deep knockoffs~\citep{Romano2019Deep}, and KnockoffGAN~\citep{Jordon2019Knockoffgan}, allow for much more general designs.
Hence, the standard knockoffs can be readily swapped for those procedures in our pipeline (without any changes to the methodology or theory) when indicated in an~application.

%An in-depth comparison with that proposal would be of interest for future research. 
Considerably later than our paper has been put online, two other papers on the topic have also been put online~\citep{Srinivasan851337,Nguyen2020Aggregation}---apparently without being aware of our manuscript. A comparison to those results would also be of interest.

%\subsubsection*{Acknowledgements}
%We thank Aaditya Ramdas and Sophie-Charlotte Klose for their valuable input.

%%%%%%%%%%%%%%%%%%%%%%%%%%%%%%%%%%%%%%%%%%
%\section{Discussion}
%
%Authors should discuss the results and how they can be interpreted from the perspective of previous studies and of the working hypotheses. The findings and their implications should be discussed in the broadest context possible. Future research directions may also be highlighted.

%%%%%%%%%%%%%%%%%%%%%%%%%%%%%%%%%%%%%%%%%%
\vspace{6pt} 

%%%%%%%%%%%%%%%%%%%%%%%%%%%%%%%%%%%%%%%%%%
%% optional
%\supplementary{The following are available online at \linksupplementary{s1}, Figure S1: title, Table S1: title, Video S1: title.}


%%%%%%%%%%%%%%%%%%%%%%%%%%%%%%%%%%%%%%%%%%
\authorcontributions{Conceptualization, F.X. and J.L.; methodology, F.X. and J.L.; software, F.X.; validation, F.X. and J.L.; writing---original draft preparation, F.X.; writing---review and editing, J.L.. All authors have read and agreed to the published version of the manuscript.}
	%For research articles with several authors, a short paragraph specifying their individual contributions must be provided. The following statements should be used ``Conceptualization, X.X. and Y.Y.; methodology, X.X.; software, X.X.; validation, X.X., Y.Y. and Z.Z.; formal analysis, X.X.; investigation, X.X.; resources, X.X.; data curation, X.X.; writing---original draft preparation, X.X.; writing---review and editing, X.X.; visualization, X.X.; supervision, X.X.; project administration, X.X.; funding acquisition, Y.Y. All authors have read and agreed to the published version of the manuscript.'', please turn to the  \href{http://img.mdpi.org/data/contributor-role-instruction.pdf}{CRediT taxonomy} for the term explanation. Authorship must be limited to those who have contributed substantially to the work~reported.}

\funding{This research received no external funding.}%Please add: ``This research received no external funding'' or ``This research was funded by NAME OF FUNDER grant number XXX.'' and  and ``The APC was funded by XXX''. Check carefully that the details given are accurate and use the standard spelling of funding agency names at \url{https://search.crossref.org/funding}, any errors may affect your future funding.}

\institutionalreview{Not applicable.}
	%In this section, please add the Institutional Review Board Statement and approval number for studies involving humans or animals. Please note that the Editorial Office might ask you for further information. Please add ``The study was conducted according to the guidelines of the Declaration of Helsinki, and approved by the Institutional Review Board (or Ethics Committee) of NAME OF INSTITUTE (protocol code XXX and date of approval).'' OR ``Ethical review and approval were waived for this study, due to REASON (please provide a detailed justification).'' OR ``Not applicable'' for studies not involving humans or animals. You might also choose to exclude this statement if the study did not involve humans or animals.}

\informedconsent{Not applicable.}
	%Any research article describing a study involving humans should contain this statement. Please add ``Informed consent was obtained from all subjects involved in the study.'' OR ``Patient consent was waived due to REASON (please provide a detailed justification).'' OR ``Not applicable'' for studies not involving humans. You might also choose to exclude this statement if the study did not involve humans.

%Written informed consent for publication must be obtained from participating patients who can be identified (including by the patients themselves). Please state ``Written informed consent has been obtained from the patient(s) to publish this paper'' if applicable.}

\dataavailability{All the software, simulations, and data analyses are provided at \href{https://github.com/LedererLab/aggregated_knockoffs}{github.com\linebreak/lederlab}. 
	The data for our analysis are downloaded from the website of the \href{http://americangut.org}{American Gut Project}, which is public.}
	%In this section, please provide details regarding where data supporting reported results can be found, including links to publicly archived datasets analyzed or generated during the study. Please refer to suggested Data Availability Statements in section ``MDPI Research Data Policies'' at \url{https://www.mdpi.com/ethics}. You might choose to exclude this statement if the study did not report any data.} 

\acknowledgments{We thank Aaditya Ramdas and Sophie-Charlotte Klose for their valuable input.}

\conflictsofinterest{The authors declare no conflict of interest.}%Declare conflicts of interest or state ``The authors declare no conflict of interest.'' Authors must identify and declare any personal circumstances or interest that may be perceived as inappropriately influencing the representation or interpretation of reported research results. Any role of the funders in the design of the study; in the collection, analyses or interpretation of data; in the writing of the manuscript, or in the decision to publish the results must be declared in this section. If there is no role, please state ``The funders had no role in the design of the study; in the collection, analyses, or interpretation of data; in the writing of the manuscript, or in the decision to publish the~results''.} 

%% Optional
%\sampleavailability{Samples of the compounds ... are available from the authors.}

%%%%%%%%%%%%%%%%%%%%%%%%%%%%%%%%%%%%%%%%%%
%% Only for journal Encyclopedia
%\entrylink{The Link to this entry published on the encyclopedia platform.}

%%%%%%%%%%%%%%%%%%%%%%%%%%%%%%%%%%%%%%%%%%
%% Optional
%\abbreviations{The following abbreviations are used in this manuscript:\\
%
%\noindent 
%\begin{tabular}{@{}ll}
%MDPI & Multidisciplinary Digital Publishing Institute\\
%DOAJ & Directory of open access journals\\
%TLA & Three letter acronym\\
%LD & Linear dichroism
%\end{tabular}}

%%%%%%%%%%%%%%%%%%%%%%%%%%%%%%%%%%%%%%%%%%
%% Optional
\appendixtitles{yes} % Leave argument "no" if all appendix headings stay EMPTY (then no dot is printed after "Appendix A"). If the appendix sections contain a heading then change the argument to "yes".
\appendixstart

\appendix
%\section{}
%\subsection{}
%The appendix is an optional section that can contain details and data supplemental to the main text---for example, explanations of experimental details that would disrupt the flow of the main text but nonetheless remain crucial to understanding and reproducing the research shown; figures of replicates for experiments of which representative data are shown in the main text can be added here if brief, or as Supplementary Data. Mathematical proofs of results not central to the paper can be added as an appendix.

\section{Additional Explanations}
\label{Appendix:additional}

\subsection{Further Simulations for Comparison to Multiple Knockoffs (MKO)~\citep{Holden2018Multiple}}\label{Appendix:subsec:MKO}
We compare the three methods---KO, AKO, and MKO---in the logistic cases. 
The results shown in Figure~\ref{fig:logistic} indicate that the MKO is more conservative than both our pipeline and the KO, and our pipeline always has higher power than KO and MKO.

%\begin{figure}[ht]
%	\centering
%	\subfigure[$(\samplesize,\numvar)=(200,100)$]{
%%		\label{fig:linear=200}
%		\begin{minipage}{0.3\textwidth}
%			\includegraphics[width=\textwidth]{figures/Conf2/linear/FDR1.pdf}
%		\end{minipage}
%		\begin{minipage}{0.3\textwidth}
%			\includegraphics[width=\textwidth]{figures/Conf2/linear/Power1.pdf}
%		\end{minipage}
%	}
%	\subfigure[$(\samplesize,\numvar)=(400,200)$]{
%%		\label{fig:linear=400}
%		\begin{minipage}{0.3\textwidth}
%			\includegraphics[width=\textwidth]{figures/Conf2/linear/FDR2.pdf}
%		\end{minipage}
%		\begin{minipage}{0.3\textwidth}
%			\includegraphics[width=\textwidth]{figures/Conf2/linear/Power2.pdf}
%		\end{minipage}
%	}
%	\caption{Our approach AKO (solid, orange circles) has a similar FDR to the standard KO (hollow, purple circles) but has more power. The MKO (solid, blue square) is more conservative than our AKO, has lower power.}%
%	\label{fig:linear}
%\end{figure}

\begin{figure}[H]
%	\centering
	\subfigure[$(\samplesize,\numvar)=(200,100)$]{
		%		\label{fig:logisticn=200}
		\begin{minipage}{0.25\textwidth}
			\includegraphics[width=\textwidth]{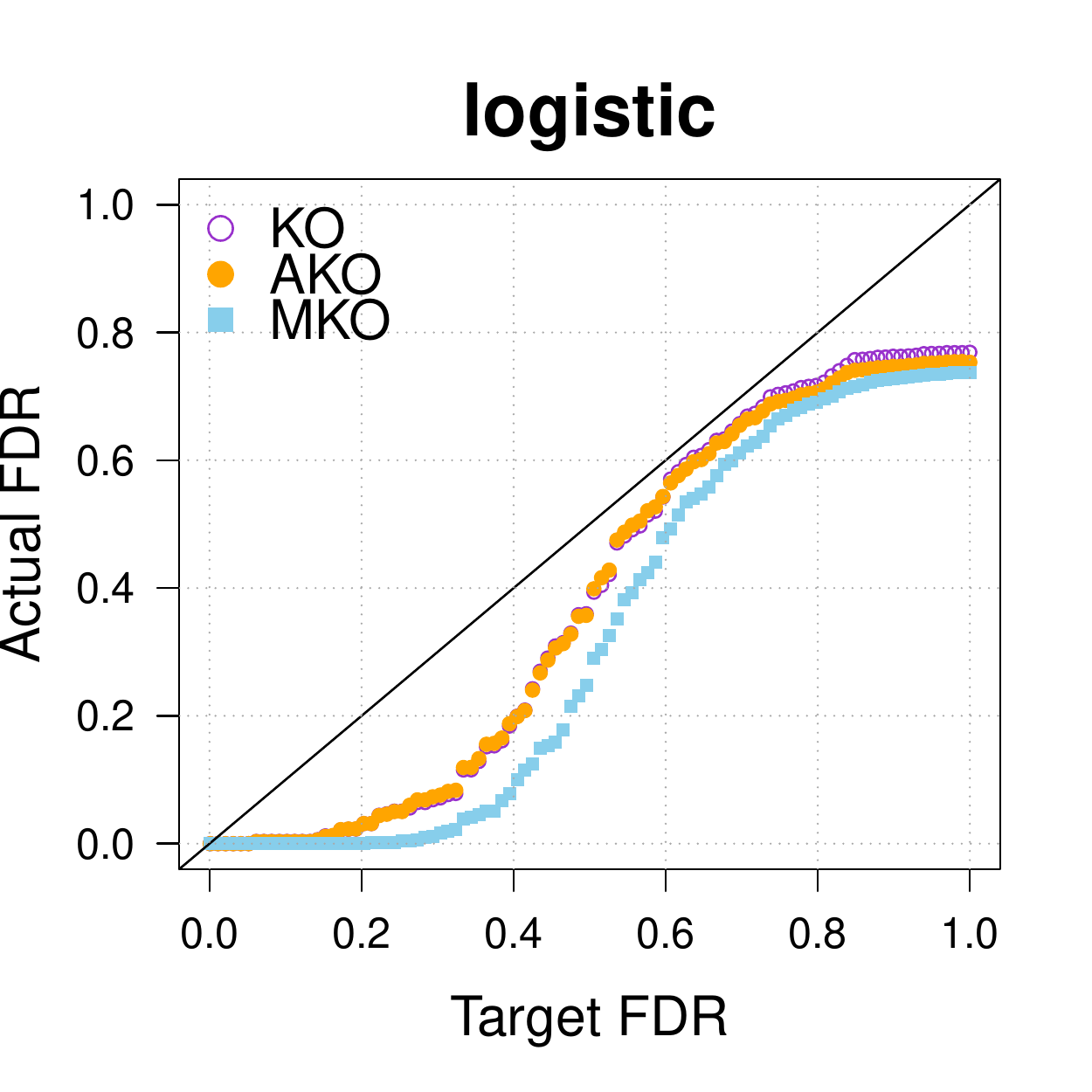}
		\end{minipage}
		\begin{minipage}{0.25\textwidth}
			\includegraphics[width=\textwidth]{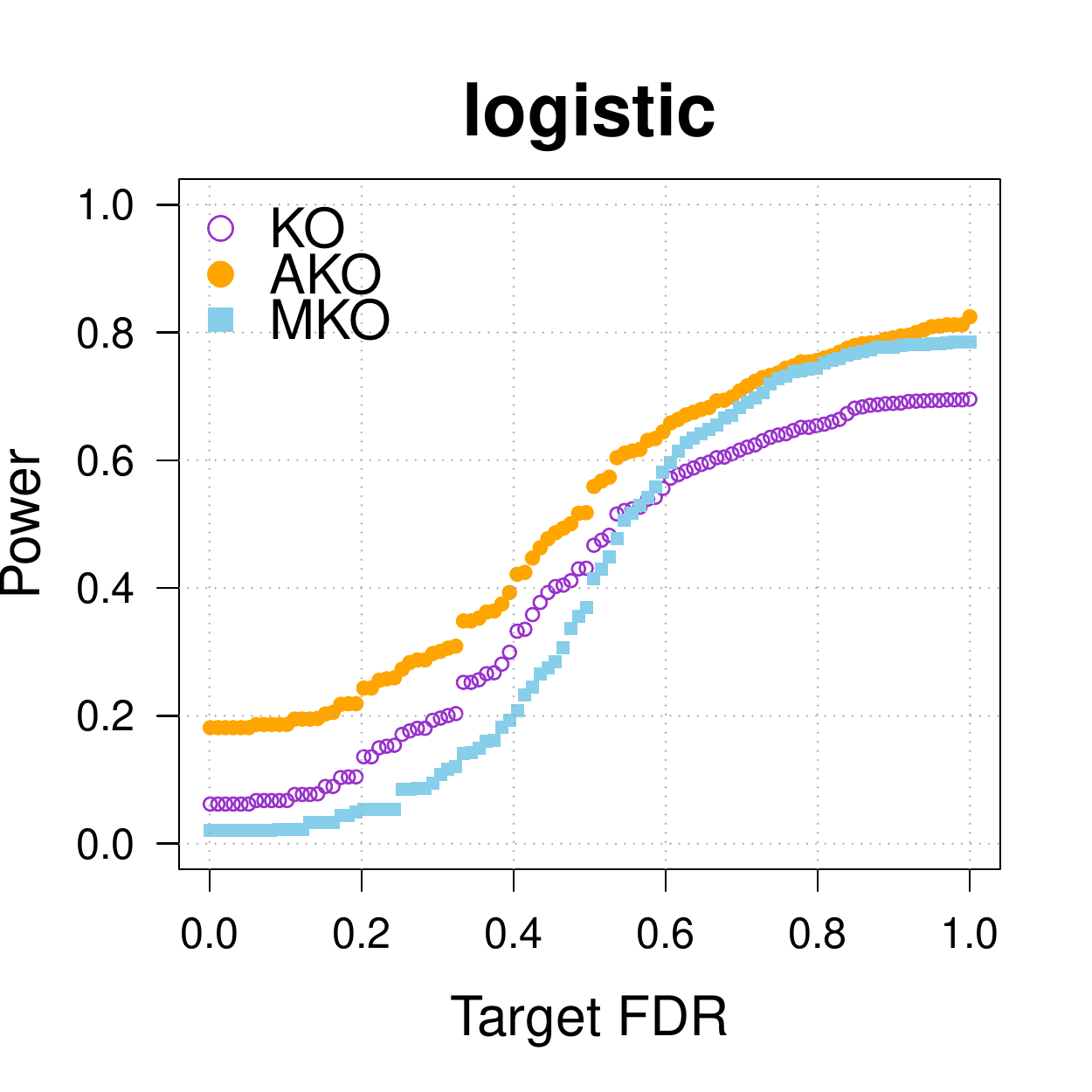}
		\end{minipage}
	}
	\subfigure[$(\samplesize,\numvar)=(400,200)$]{
		%		\label{fig:logisticn=400}
		\begin{minipage}{0.25\textwidth}
			\includegraphics[width=\textwidth]{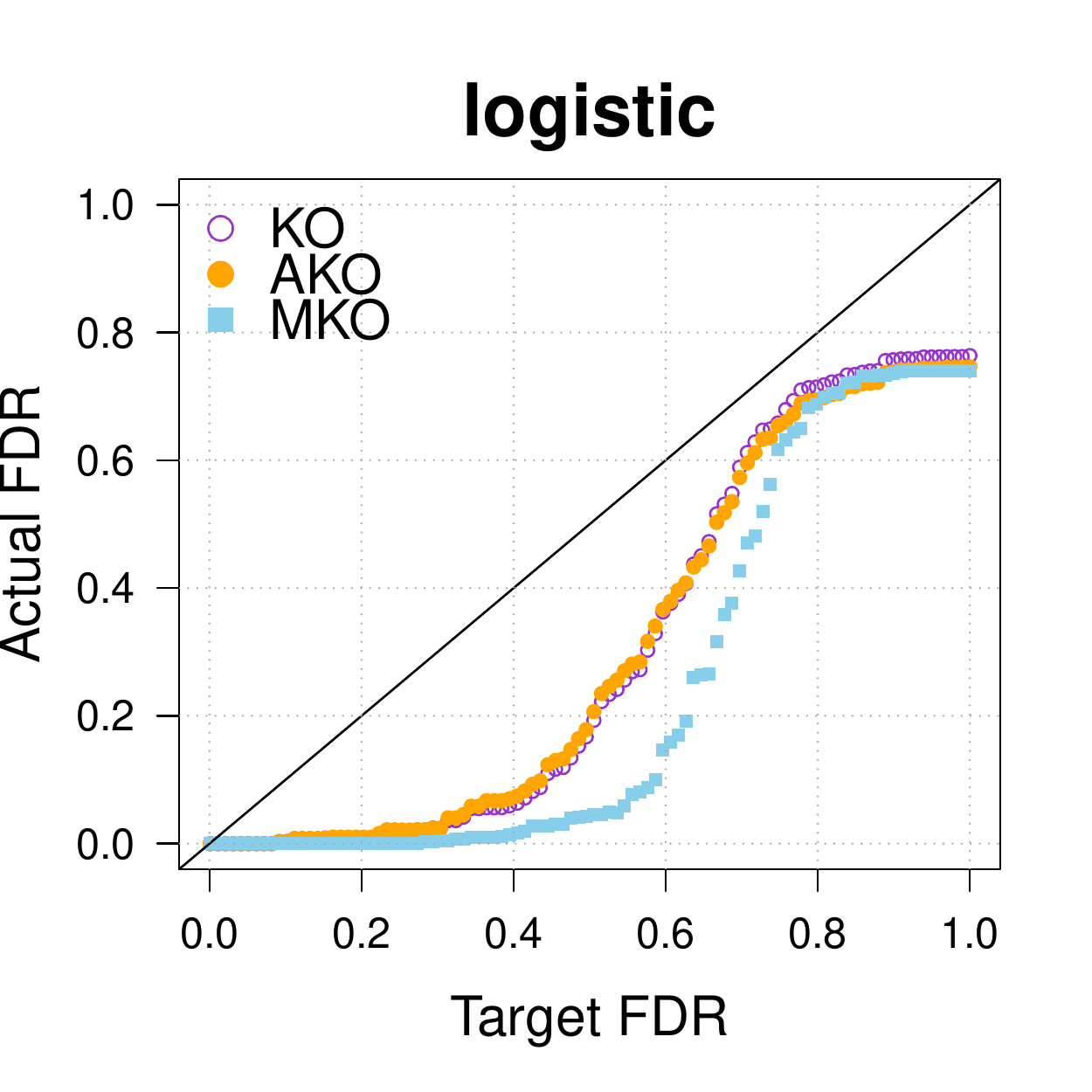}
		\end{minipage}
		\begin{minipage}{0.25\textwidth}
			\includegraphics[width=\textwidth]{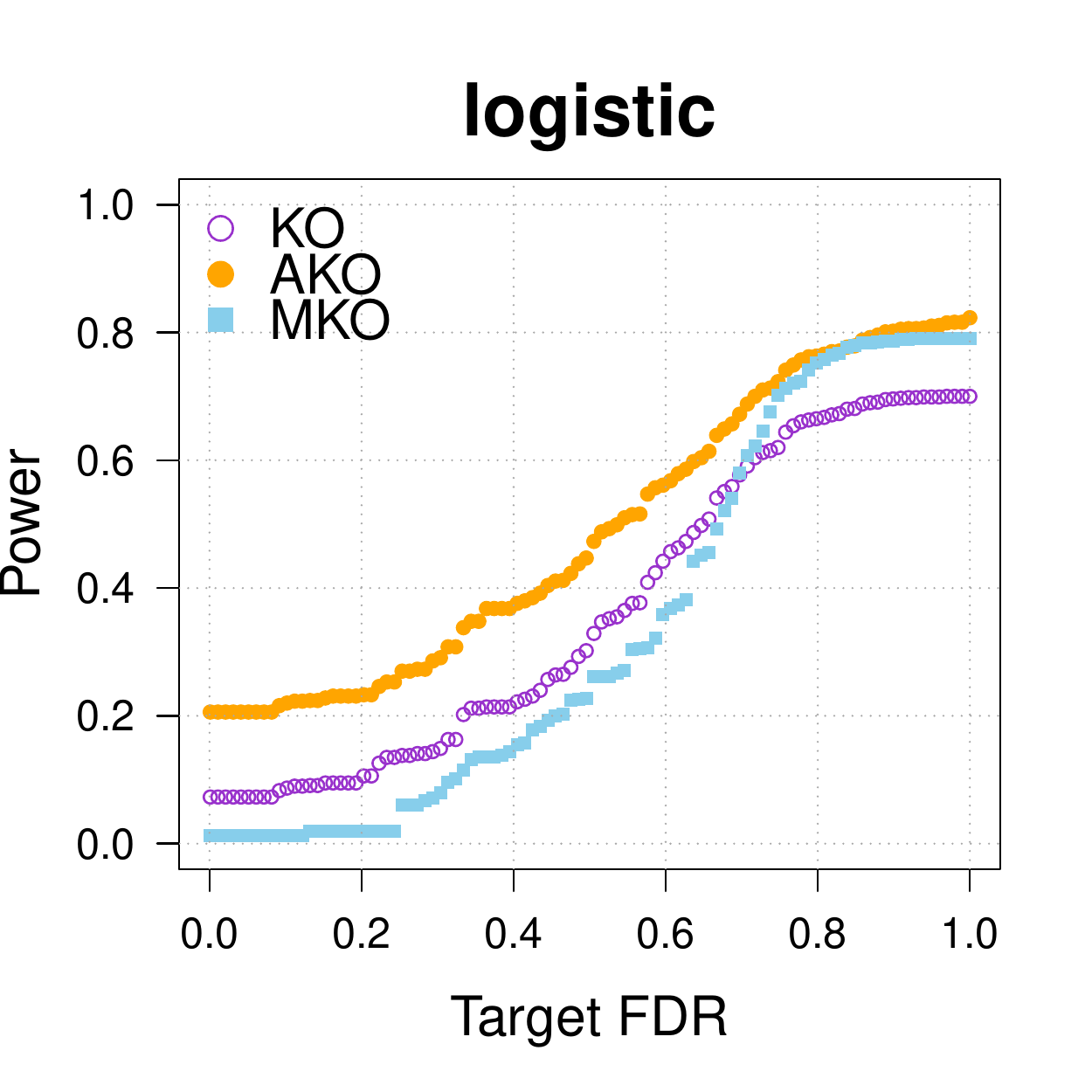}
		\end{minipage}
	}
	\caption{Our approach AKO (solid, orange circles) has a similar FDR to the standard KO (hollow, purple circles) but has more power. The MKO (solid, blue square) is more conservative than our AKO, has lower power.}%
	\label{fig:logistic}
\end{figure}

\subsection{Choice of $q_1,\dots,q_k$}\label{Appendix:subsec:choice-of-qi}
The theory applies to every choice of $q_1,\dots,q_k$ that satisfies the condition in step 1 on page~\pageref{choice-of-qi}, but indeed, each choice has slightly different characteristics in practice.
To give an example,
we have added the result for $q_i=q/k$ (AKO.ave, Figure~\ref{fig:ave}) to Figure~\ref{fig1}a in the main text.
We observe in general that the results do look different for each choice of the sequence but that the differences are usually only small.%\vspace{-4mm}
\begin{figure}[H]
%	\centering
	%	\vspace{-2mm}
	\includegraphics[width=0.25\textwidth]{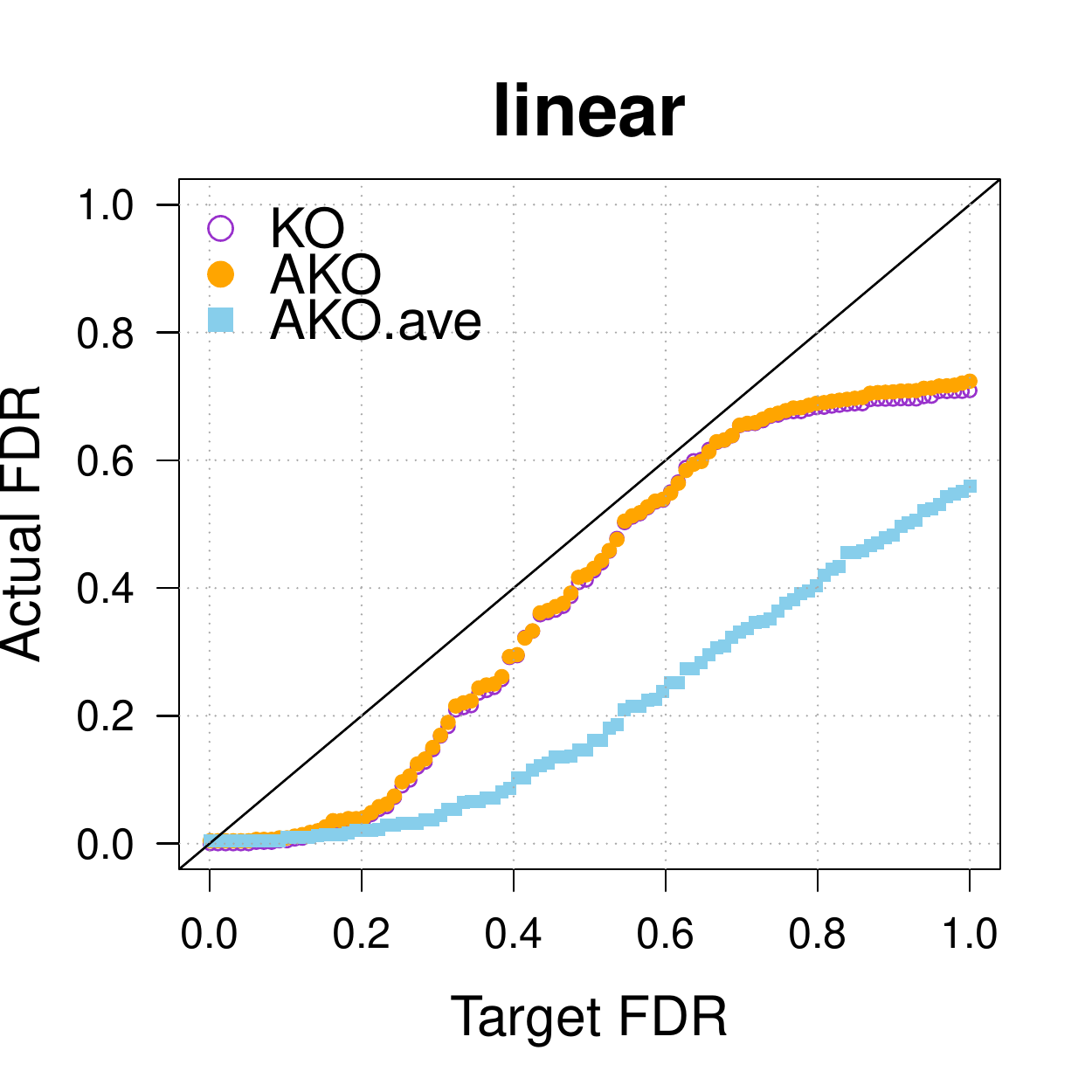}
	\includegraphics[width=0.25\textwidth]{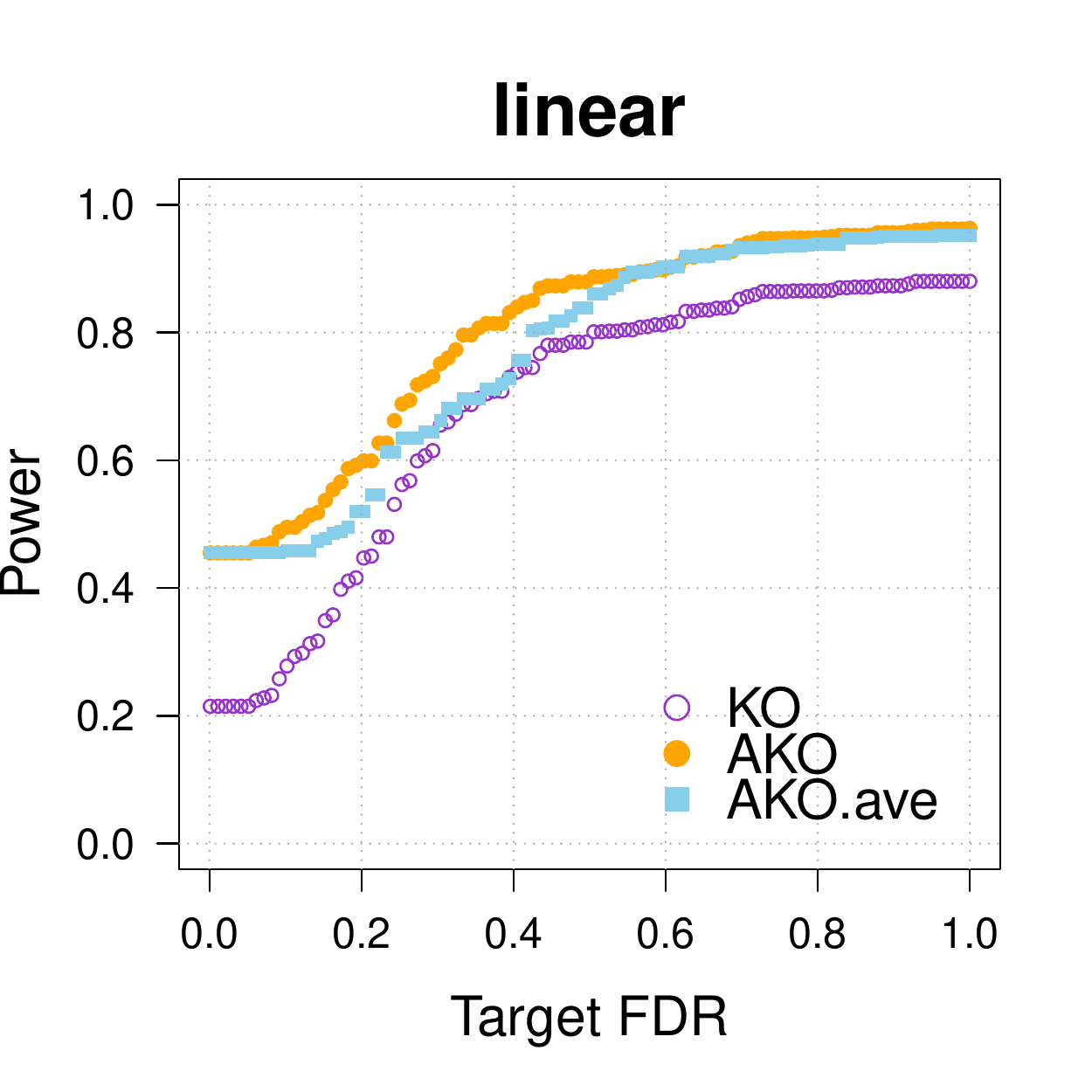}
	\caption{Actual FDR and power of the KO, AKO, AKO.ave in the linear case with $(\samplesize,\numvar)=(200,100)$. }\label{fig:ave}
	%The plots correspond to Figure~1(a) on Page~5 in the paper.}\label{fig:qi}%
	%	\vspace{-0.2cm}
\end{figure}

\subsection{Various Settings for the Simulation Part}\label{Appendix:subsec:various-settings}
Our conclusions hold over a wide spectrum of settings,
including different dimensionalities, 
sparsity levels, correlations, and so forth.
To illustrate this,
we vary the settings of the linear case (see Figure~\ref{fig1}a,b) once more in Figure~\ref{fig:two} below.%\vspace{-4mm}

\begin{figure}[ht]
%	\centering
	%	\vspace{-2mm}
	\subfigure[$(\samplesize,\numvar,s,\rho)=(300,100,30,0.5)$]{
		\begin{minipage}{0.3\textwidth}
			\includegraphics[width=\textwidth]{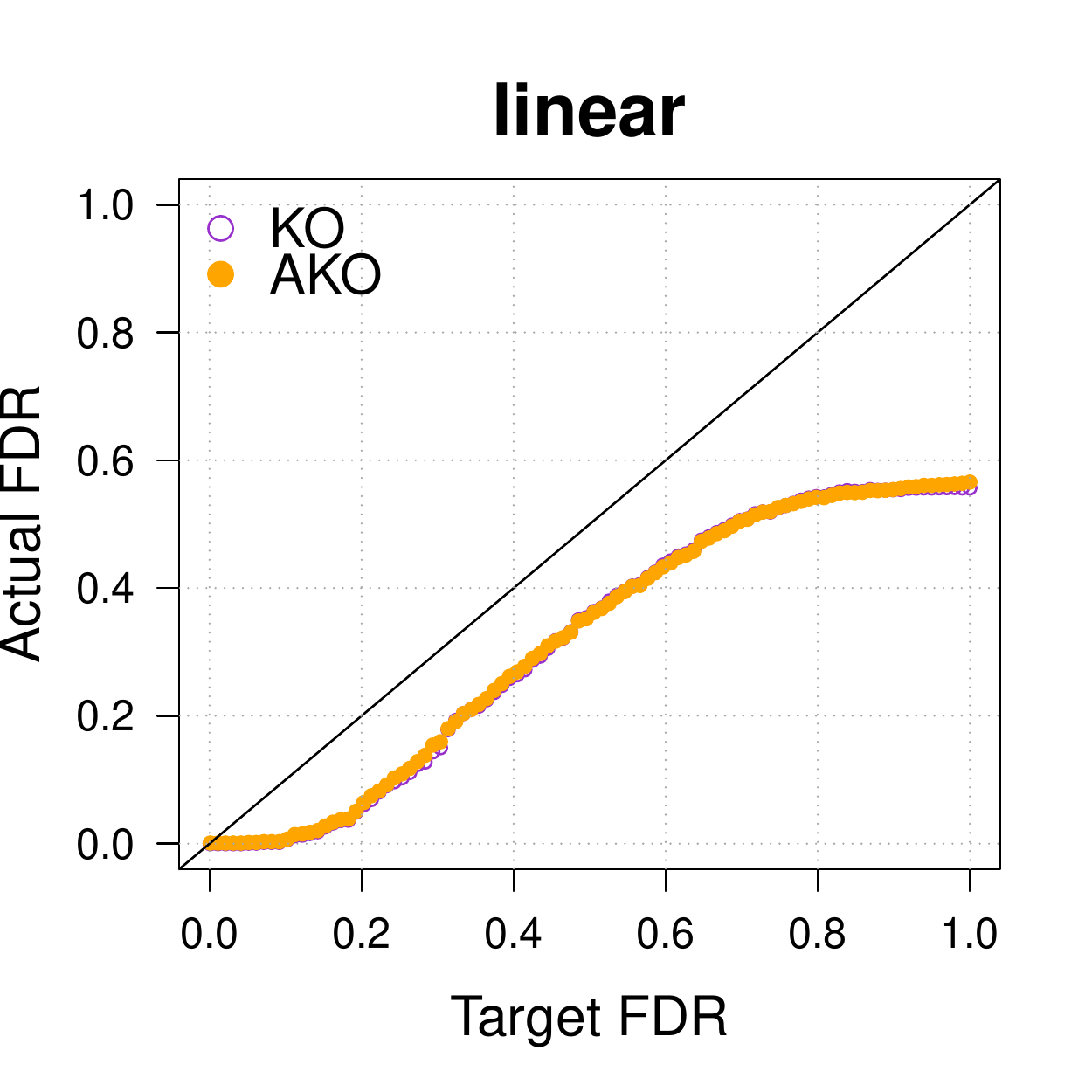}
		\end{minipage}
		\begin{minipage}{0.3\textwidth}
			\includegraphics[width=\textwidth]{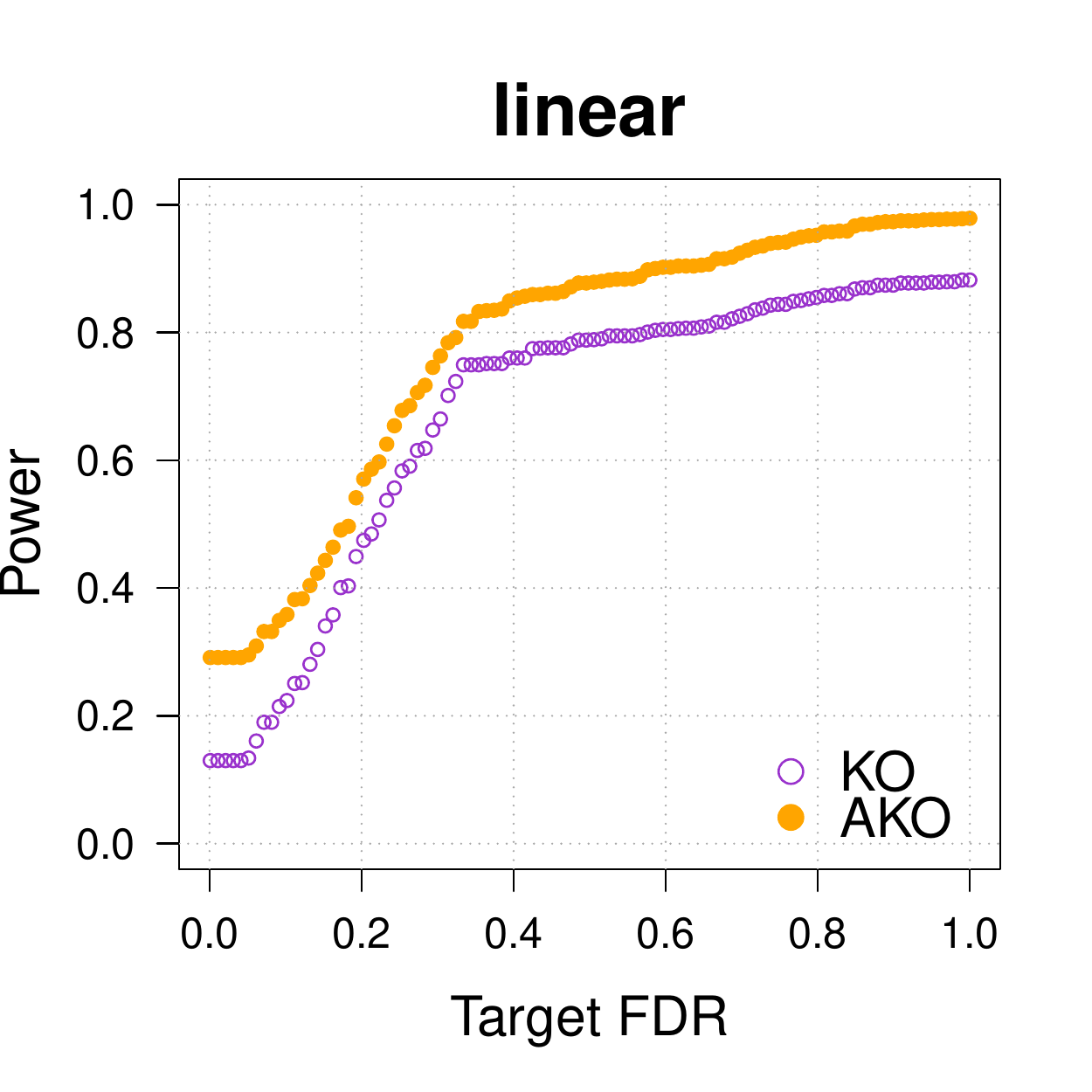}
	\end{minipage}}
	
	\subfigure[$(\samplesize,\numvar,s,\rho)=(500,200,40,0.5)$]{
		\begin{minipage}{0.3\textwidth}
			\includegraphics[width=\textwidth]{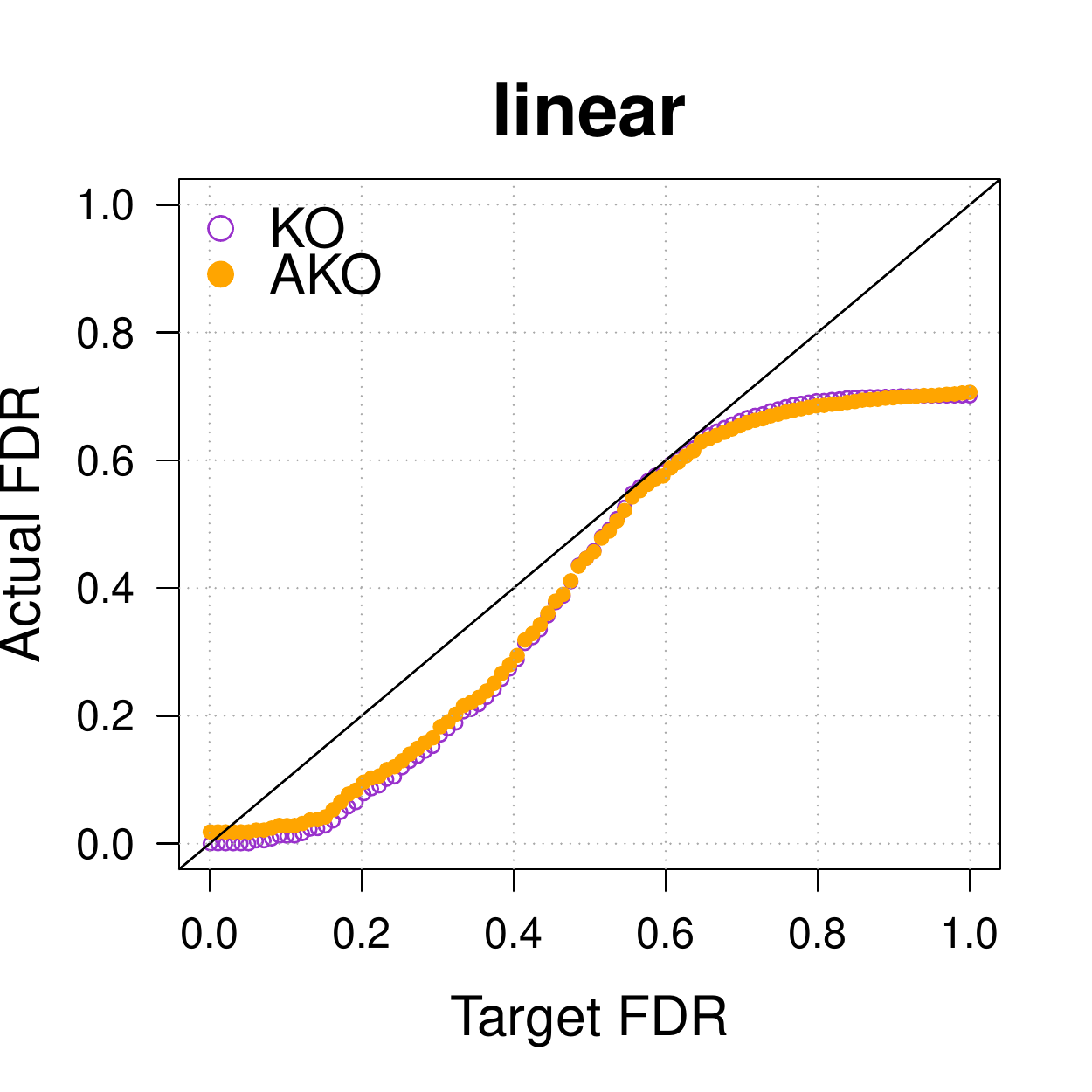}
		\end{minipage}
		\begin{minipage}{0.3\textwidth}
			\includegraphics[width=\textwidth]{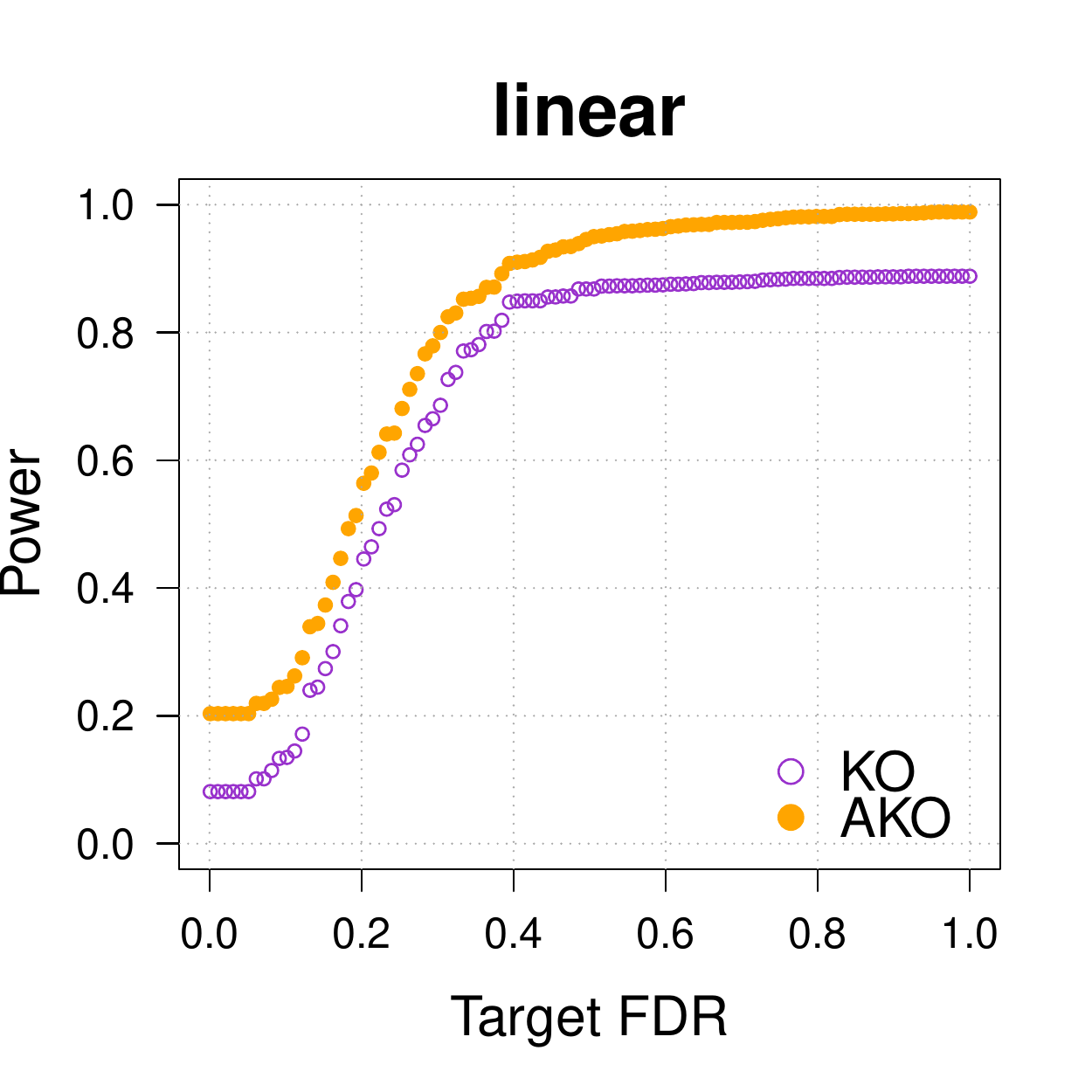}
	\end{minipage}}
	\caption{Our approach AKO (solid, orange circles) has a similar FDR to the standard KO (hollow, purple circles) but has more power.}%
	\label{fig:two}
\end{figure}

\subsection{Better than other Competitors (under the AGP Data)}\label{Appendix:subsec:competitors}
Knockoff methods seem to be better suited for the microbiome data than their competitors.
For illustration,
we have included  BH~\citep{BH95} and TreeFDR~\citep{Xiao2017False} in the application that  corresponds to Table~\ref{tab:realdatalogistic}~(i) in the paper.
We find (in Table~\ref{tab:Tree-BH}) on those data more generally that the knockoff methods select many more predictors than other methods.

\begin{specialtable}[H] %MDPI: Is the italics necessary?
	\caption{{Selected bacterial phyla by four methods}---BH, TreeFDR, KO, and AKO (correponds to Table~\ref{tab:realdatalogistic} (i)).}\label{tab:Tree-BH}
	\centering
%	\scriptsize
\setlength{\cellWidtha}{\columnwidth/4-2\tabcolsep+0in}
\setlength{\cellWidthb}{\columnwidth/4-2\tabcolsep+0in}
\setlength{\cellWidthc}{\columnwidth/4-2\tabcolsep-0in}
\setlength{\cellWidthd}{\columnwidth/4-2\tabcolsep-0in}
\scalebox{1}[1]{\begin{tabularx}{\columnwidth}{>{\PreserveBackslash\centering}p{\cellWidtha}>{\PreserveBackslash\centering}p{\cellWidthb}>{\PreserveBackslash\centering}p{\cellWidthc}>{\PreserveBackslash\centering}p{\cellWidthd}}
\toprule

		\multicolumn{4}{c}{\textbf{(i) ALL}} \\
		\textbf{BH} & \textbf{TreeFDR} & \textbf{KO} &  \textbf{AKO} \\
		\hline
		&& \emph{Actinobacteria} & \emph{Actinobacteria} \\
		&&& \emph{Bacteroidetes}  \\ 
		& \emph{Cyanobacteria} & \emph{Cyanobacteria} & \emph{Cyanobacteria} \\
		&& \emph{Proteobacteria} & \emph{Proteobacteria} \\ 
		&&&  \emph{Spirochaetes}  \\ 
		&& \emph{Synergistetes} & \emph{Synergistetes}  \\
		&& \emph{Tenericutes} & \emph{Tenericutes} \\
		\emph{Verrucomicrobia} & \emph{Verrucomicrobia} & & \emph{Verrucomicrobia}  \\ 
		\bottomrule
	\end{tabularx}}
	\label{fig:tab}
\end{specialtable}

\section{Additional Results on the Genera Rank}\label{Appendix:genera}
We present in this section the results on the genera rank of different groupings; see Tables~\ref{tab:genera1}--\ref{tab:genera6}.

\begin{specialtable}[H] %MDPI: Is the italics necessary?
%	\centering
	\caption{{Analysis at the genus level rank for the grouping} (ii) uw+ob.
		AKO selects more genera than the original KO.}\label{tab:genera1}
	\setlength{\cellWidtha}{\columnwidth/3-2\tabcolsep+0in}
\setlength{\cellWidthb}{\columnwidth/3-2\tabcolsep+0in}
\setlength{\cellWidthc}{\columnwidth/3-2\tabcolsep-0in}
\scalebox{1}[1]{\begin{tabularx}{\columnwidth}{>{\PreserveBackslash\centering}p{\cellWidtha}>{\PreserveBackslash\centering}p{\cellWidthb}>{\PreserveBackslash\centering}p{\cellWidthc}}
\toprule
		\textbf{Phylum} & \textbf{KO} & \textbf{AKO}\\
		\hline 
		Actinobacteria & Collinsella & Collinsella \\ 
		\hline
		\multirow{3}{*}{Firmicutes} 
		& & Lachnospira \\
		& & Acidaminococcus \\
		& & Catenibacterium\\
		\hline
		Tenericutes & RF39 & RF39 \\
		\bottomrule
	\end{tabularx}}
\end{specialtable}

\begin{specialtable}[H] 
	\centering
	\caption{Analysis at the genus level for the grouping (iii) nor + ob.
		AKO selects more genera than the original KO.}\label{tab:genera2}
	\setlength{\cellWidtha}{\columnwidth/3-2\tabcolsep+0in}
\setlength{\cellWidthb}{\columnwidth/3-2\tabcolsep+0in}
\setlength{\cellWidthc}{\columnwidth/3-2\tabcolsep-0in}
\scalebox{1}[1]{\begin{tabularx}{\columnwidth}{>{\PreserveBackslash\centering}p{\cellWidtha}>{\PreserveBackslash\centering}p{\cellWidthb}>{\PreserveBackslash\centering}p{\cellWidthc}}
\toprule

		\textbf{Phylum} & \textbf{KO} & \textbf{AKO}\\
		\hline 
		\multirow{2}{*}{Actinobacteria} & Actinomyces & Actinomyces\\
		& Collinsella & Collinsella \\ 
		\hline  
		Cyanobacteria & YS2 & YS2 \\ 
		\hline
		\multirow{9}{*}{Firmicutes} & Bacillus & Bacillus \\
		& & Lactococcus \\
		& Lachnospira & Lachnospira \\
		& Ruminococcus & Ruminococcus \\ 
		& Acidaminococcus & Acidaminococcus \\
		& Megasphaera & Megasphaera \\
		& & Mogibacteriaceae \\
		& & Erysipelotrichaceae \\ 
		& Catenibacterium & Catenibacterium\\
		\hline
		\multirow{2}{*}{Proteobacteria} & RF32 & RF32 \\ 
		& & Haemophilus \\
		\hline
		\multirow{2}{*}{Tenericutes} & RF39 & RF39 \\
		& & ML615J-28 \\
		\bottomrule
	\end{tabularx}}
\end{specialtable}

\begin{specialtable}[H] 
	\centering
	\caption{Analysis at the genus level for the grouping (iv) ow + ob.
		AKO selects more genera than the original KO.}\label{tab:genera3}
	\setlength{\cellWidtha}{\columnwidth/3-2\tabcolsep+0in}
\setlength{\cellWidthb}{\columnwidth/3-2\tabcolsep+0in}
\setlength{\cellWidthc}{\columnwidth/3-2\tabcolsep-0in}
\scalebox{1}[1]{\begin{tabularx}{\columnwidth}{>{\PreserveBackslash\centering}p{\cellWidtha}>{\PreserveBackslash\centering}p{\cellWidthb}>{\PreserveBackslash\centering}p{\cellWidthc}}
\toprule

		\textbf{Phylum} & \textbf{KO} & \textbf{AKO}\\
		\hline 
		Actinobacteria 
		& Eggerthella & Eggerthella \\
		\hline  
		\multirow{2}{*}{Cyanobacteria} & YS2 & YS2 \\ 
		& Streptophyta & Streptophyta \\
		\hline
		\multirow{7}{*}{Firmicutes} 
		& & Bacillus \\
		& Clostridium & Clostridium \\
		& Lachnospira & Lachnospira \\
		& Acidaminococcus & Acidaminococcus \\
		& & 1-68 \\
		& Erysipelotrichaceae & Erysipelotrichaceae \\ 
		& & Catenibacterium\\
		\hline
		Proteobacteria
		& Haemophilus & Haemophilus \\
		\bottomrule
	\end{tabularx}}
\end{specialtable}

\begin{specialtable}[H] 
	\centering
	\caption{Analysis at the genus level for the grouping (v) uw + nor + ob.
		AKO selects more genera than the original KO.}\label{tab:genera4}
	\setlength{\cellWidtha}{\columnwidth/3-2\tabcolsep+0in}
\setlength{\cellWidthb}{\columnwidth/3-2\tabcolsep+0in}
\setlength{\cellWidthc}{\columnwidth/3-2\tabcolsep-0in}
\scalebox{1}[1]{\begin{tabularx}{\columnwidth}{>{\PreserveBackslash\centering}p{\cellWidtha}>{\PreserveBackslash\centering}p{\cellWidthb}>{\PreserveBackslash\centering}p{\cellWidthc}}
\toprule

		\textbf{Phylum} & \textbf{KO} & \textbf{AKO}\\
		\hline 
		\multirow{2}{*}{Actinobacteria} & Actinomyces & Actinomyces\\
		& Collinsella & Collinsella \\ 
		\hline  
		Cyanobacteria & YS2 & YS2 \\
		\hline
		\multirow{10}{*}{Firmicutes} & Bacillus & Bacillus \\
		& & Lactococcus \\
		& Lachnospira & Lachnospira \\
		& Ruminococcus & Ruminococcus \\ 
		& Acidaminococcus & Acidaminococcus \\
		& Megasphaera & Megasphaera \\
		& & Mogibacteriaceae \\
		& & SHA-98 \\
		& & Erysipelotrichaceae \\ 
		& Catenibacterium & Catenibacterium\\
		\hline
		\multirow{2}{*}{Proteobacteria} & RF32 & RF32 \\ 
		& & Haemophilus \\
		\hline
		\multirow{2}{*}{Tenericutes} & RF39 & RF39 \\
		& & ML615J-28 \\
		\bottomrule
	\end{tabularx}}
\end{specialtable}

\begin{specialtable}[H] 
	\centering
	\caption{Analysis at the genus level for the grouping (vi) uw + ow + ob.
		AKO selects more genera than the original KO.}\label{tab:genera5}
	\setlength{\cellWidtha}{\columnwidth/3-2\tabcolsep+0in}
\setlength{\cellWidthb}{\columnwidth/3-2\tabcolsep+0in}
\setlength{\cellWidthc}{\columnwidth/3-2\tabcolsep-0in}
\scalebox{1}[1]{\begin{tabularx}{\columnwidth}{>{\PreserveBackslash\centering}p{\cellWidtha}>{\PreserveBackslash\centering}p{\cellWidthb}>{\PreserveBackslash\centering}p{\cellWidthc}}
\toprule
		\textbf{Phylum} & \textbf{KO} & \textbf{AKO}\\
		\hline 
		Actinobacteria
		& Eggerthella & Eggerthella \\
		\hline  
		\multirow{2}{*}{Cyanobacteria} & YS2 & YS2 \\ 
		& Streptophyta & Streptophyta \\
		\hline
		\multirow{10}{*}{Firmicutes} 
		& Bacillus & Bacillus \\
		& & Lactobacillus \\ 
		& Clostridium & Clostridium \\
		& Lachnospira & Lachnospira \\
		& & Veillonellaceaes \\ 
		& Acidaminococcus & Acidaminococcus \\
		& 1-68 & 1-68 \\
		& Erysipelotrichaceae & Erysipelotrichaceae \\ 
		& Catenibacterium & Catenibacterium\\
		& Eubacterium & Eubacterium \\
		\hline
		\multirow{2}{*}{Proteobacteria} & & RF32 \\ 
		& Haemophilus & Haemophilus \\
		\bottomrule
	\end{tabularx}}
\end{specialtable}

\begin{specialtable}[H] 
	\centering
	\caption{Analysis at the genus level for the grouping (vii) nor+ow+ob.
		AKO selects more genera than the original KO.}\label{tab:genera6}
	\setlength{\cellWidtha}{\columnwidth/3-2\tabcolsep+0in}
\setlength{\cellWidthb}{\columnwidth/3-2\tabcolsep+0in}
\setlength{\cellWidthc}{\columnwidth/3-2\tabcolsep-0in}
\scalebox{1}[1]{\begin{tabularx}{\columnwidth}{>{\PreserveBackslash\centering}p{\cellWidtha}>{\PreserveBackslash\centering}p{\cellWidthb}>{\PreserveBackslash\centering}p{\cellWidthc}}
\toprule

		\textbf{Phylum} & \textbf{KO} & \textbf{AKO}\\
		\hline 
		\multirow{3}{*}{Actinobacteria} & Actinomyces & Actinomyces\\
		& Collinsella & Collinsella \\ 
		& Eggerthella & Eggerthella \\
		\hline  
		Cyanobacteria & YS2 & YS2 \\ 
		\hline
		\multirow{7}{*}{Firmicutes} & Bacillus & Bacillus \\
		& Lachnospira & Lachnospira \\
		& Ruminococcus & Ruminococcus \\
		& Acidaminococcus & Acidaminococcus \\
		& Megasphaera & Megasphaera \\
		& Erysipelotrichaceae & Erysipelotrichaceae \\ 
		& Catenibacterium & Catenibacterium\\
		\hline
		\multirow{2}{*}{Proteobacteria} & RF32 & RF32 \\ 
		& Haemophilus & Haemophilus \\
		\hline
		Tenericutes & & RF39 \\
		\bottomrule
	\end{tabularx}}
\end{specialtable}

%\section{}
%All appendix sections must be cited in the main text. In the appendices, Figures, Tables, etc. should be labeled, starting with ``A''---e.g., Figure A1, Figure A2, etc. 

%%%%%%%%%%%%%%%%%%%%%%%%%%%%%%%%%%%%%%%%%%
\end{paracol}
\reftitle{References}

\end{document}